\documentclass[11pt,a4paper]{article}
\usepackage{amssymb} \usepackage{amsmath} \usepackage{graphicx}
\usepackage{epsfig,latexsym,color,xcolor}
\usepackage{ulem}
\usepackage{amstext}    
\usepackage{array} 
\usepackage{slashed}
\usepackage[height=9.5in ,hmargin={1.8cm,0.8in}]{geometry}
\usepackage{subfig}
\usepackage{booktabs}
\newcommand{\RNum}[1]{\uppercase\expandafter{\romannumeral #1\relax}}
\usepackage{braket}

\DeclareGraphicsExtensions{.png, .pdf, .eps,}
\newcommand{\p}{\psi}

\newcommand{\al}{\alpha'}

\newcommand{\pa}{\partial}

\makeatletter
\newcommand*{\rom}[1]{\expandafter\@slowromancap\romannumeral #1@}
\makeatother

\numberwithin{equation}{section}

\begin{document}

\large
 \begin{center}
 {\Large \bf   String coherent vertex operators of Neveu-Schwarz and Ramond states}

 \end{center}

 \vspace{0.1cm}
 \begin{center}
{\large Alice Aldi\footnote{E-mail: alice.aldi@roma2.infn.it}
 and Maurizio Firrotta\footnote{E-mail: maurizio.firrotta@roma2.infn.it}}

\vspace{0.5cm}
{\it Dipartimento di Fisica, Universit\`a di Roma Tor Vergata, \\
I.N.F.N. Sezione di Roma Tor Vergata, \\
Via della Ricerca Scientifica, 1 - 00133 Roma, ITALY}
\end{center}

\vspace{1cm}
\begin{abstract}
We present a compact expression for a coherent vertex operator in superstring theory, in the Neveu-Schwarz sector, that can be easily extended to the Ramond sector by supersymmetric transformations in target space. We give also an explicit construction for the Ramond coherent vertex operator. These constructions provide the use of the supersymmetric version of the Del Giudice-Di Vecchia-Fubini operators and Operator Product Expansion techniques, displaying manifest BRST invariance and after normal ordering also manifest covariance. As a consistency check we compute tree level three-points scattering amplitudes involving Neveu-Schwarz coherent vertex operators in different superghost pictures.
Extending these results to the closed superstring, we give the expressions of the coherent vertex operators in the case of type \RNum{2}A and \RNum{2}B theories.
\end{abstract}

\vspace{1cm}
\tableofcontents 
\newpage

\section*{Introduction}
 In the early days, firstly in the context of bosonic string theory  and after some time in its supersymmetric extension, a quantization technique was developed in the light-cone gauge based on the introduction of operators expressed in terms of (super)conformal primary fields. This approach reproduces exactly the (super)algebra needed to quantize the theory including the absence of unphysical states i.e. it is manifestly BRST invariant \cite{DelGiudice:1971yjh, Brower:1972wj, Brower:1973iz, Brower:1971qr, Thorn:2011ii, Kohriki:2010ry}. The complete set of these operators goes under the name of Spectrum-Generating-Algebra (SGA) composed by the (super) Del Giudice-Di Vecchia-Fubini (DDF) operators. The operators with longitudinal components reproduce the (super)Virasoro algebra while the operators with transverse components reproduce the (super)Heisenberg algebra. The application and generalization of the SGA was introduced in many context of string theory, for example to investigate the physical spectrum of charged strings (bosonic string, superstring and heterotic string) \cite{Kokado:2006ti, Kokado:2006ia, Kokado:2007fb, Kokado:2009bw, Kokado:2010by}, or in different superstring formulations, to build the corresponding vertex operators \cite{Mukhopadhyay:2005zu, Jusinskas:2014vqa, Skliros:2009cs, Skliros:2011zz, Skliros:2011si, Hindmarsh:2010if}. Some work regarding the study of the physical spectrum has been done also in the case of non critical strings, both bosonic and fermionic, and also for the massive string \cite{Jaskolski:1994jt, Hasiewicz:1995wr, Daszkiewicz:1997ax, Hasiewicz:1999nr}. The power of the DDF operators emerges also in the computation of string scattering amplitudes such as in the work \cite{Friedman:1971ni} for the representation of multistates scattering amplitudes in bosonic string, \cite{Ademollo:1974kz, DiVecchia:1986jv} for the explicit computation of three-points amplitude of arbitrarily massive states still in bosonic string and  \cite{Hornfeck:1987wt} for the generalization of the three-points amplitude in the Neveu-Schwartz model. Finally recent progress, using different contextualizations of  the DDF operators, has been done in \cite{SklirosOAC} for the decay of  macroscopic strings and in \cite{Bianchi:2019ywd} for the computation of three and four-points scattering amplitudes involving the open bosonic coherent states constructed in \cite{Skliros:2011si, Hindmarsh:2010if} finding a new saddle point in the Veneziano-like amplitudes. 

Encouraged by the many powerful aspects of the DDF operators we have chosen this formalism, firstly to derive the construction of arbitrarily massive states of the Ramond-Neveu-Schwarz open string, in particular in the Neveu-Schwarz (NS) sector, and finally construct coherent states in both NS and Ramond (S) sectors. Relying on the operator state correspondence \cite{Friedan:1985ge} one is free to move from one description to the other, as we did. To be more precise, we have considered the critical open superstring with only a single D9 brane, but the topics treated in the work remain still valid considering more than one brane, after including the Chan-Paton factors \cite{Bianchi:1990yu}. 

The purpose of constructing such vertex operators (namely, arbitrarily massive vertex operators and coherent vertex operators) is to make a bridge between the light-cone formalism, where the string spectrum is easy to handle, and the Conformal Field Theory (CFT) technology where scattering amplitude computation is often feasible. This is done building BRST invariant and fully covariant operators with a smart identification of the polarisations content (physical degrees of freedom) directly related to the mass shell.

Indeed, the possibility of generating arbitrarily massive vertex operators (or arbitrarily higher spin operators) automatically including the physical polarisations is extremely  useful to be able to explicitly compute complicated higher spin couplings. 
In addition, from the arbitrarily massive vertex operator, looking recursively at each mass level and summing all the contributions, we have found a non trivial exponentiation~\footnote{To be constrained by GSO projections.} which gives rise to a coherent vertex operator preserving all the listed properties and reproducing all the possible vertex operators at any given mass level~\footnote{There is one to one correspondence between the degeneration of the states and the representative vertex operators for each mass level.}. 

In section \ref{sec1}  will be derived the arbitrarily massive vertex operators for  the NS sector at fixed superghost picture, that for semplicity we have chosen to be the canonical one ($q=-1$). Successively we will extend the NS coherent vertex operator to the first non canonical superghost picture ($q=0$) using the picture rising operator, showing that picture changing can be easily and explicitly exploited. As a consequence all the techniques developed in \cite{Friedan:1985ge} are still valid in the case of string macroscopic state.
 For both superghost representations we will display the mass level expansion and show how to identify the combiantion of operators representing the expected operatorial and polarisation content for fixed mass level. 
 
In section 2, as a consistency check for the structure of the NS coherent vertex operator, we will compute the three-points coupling between two vectors and one coherent state, repeating the computation for both superghost pictures. 

In section 3, to conclude the discussion about coherent states in superstring theory, we will discuss how to derive the coherent vertex operator in the R sector exposing two different procedures.
Finally we will conclude with comments on how to extend the coherent states derived to closed superstring theory giving formulae for the vertex operators fot the \RNum{2}A and \RNum{2}B theories.

\section{ Supersymmetric DDF operators, Spectrum-Generating-Algebra \\ and~coherent~vertex~operators}\label{sec1}
In this section we will briefly review the basic aspects related of the superstring\footnote{to be more precise superstring theory in critical dimension.} spectrum focusing on the construction of physical states using the DDF formalism. After presenting the original setup we will discuss how to generate states, using OPE techniques and  choosing a convenient picture representation for the vacuum state. We will compute explicitly two representative OPE's in order to display the general properties which will be recurrent in the rest of the paper.
Finally we will construct coherent vertex operators for NS states in the canonical and non-canonical picture.

The starting point is the superstring mass-shell in the NS-sector
\begin{equation} \label{m1}
\alpha'\mathbb{M}^{2}= n^{(\alpha)}+{n^{(b)}/2}-1/2 \,; \quad  n^{(\alpha)} \geq 0\,, n^{(b)} \geq 0
\end{equation}
in which $n^{(\alpha)}$ and $n^{(b)}$ are integers and represent respectively the eigenvalues of the number operator of worldsheet bosons and fermions with $\alpha_n^i$ and $b_{n/2}^i$ the transverse oscillation modes.
As explained in \cite{Gliozzi:1976qd}, in order to have a tachyon free theory, in the NS sector the Gliozzi-Scherk-Olive (GSO) projection forbids half-integer mass states, as a result $n^{(b)}\,\rightarrow\, 2m^{(b)} - 1$ and the mass shell becomes

\begin{equation} \label{m2}
\alpha'\mathbb{M}^{2}= n^{(\alpha)}+{m^{(b)}}-1\,; \quad n^{(\alpha)} \geq 0\,, m^{(b)} > 0
\end{equation}

One way to reproduce exactly this mass shell and span the full physical space of states\footnote{in principle the DDF formalism is not manifestly covariant, but as we will discuss, there is at least one covariantization map.} is to consider the DDF quantization approach, that provides the spectrum generating algebra (SGA) of the system. For our purposes it is sufficient to consider only the transverse algebra of the SGA consisting of the transverse DDF operators : 
\begin{equation}\label{Addf}
{\cal A}^{i}_{n}={1\over 2\alpha'}\oint {dz \over 2\pi i}\, \left( i\pa X^{i} + 2\alpha' n \,q{\cdot}\p \,\p^{i} \right) e^{inq{\cdot X}} 
\end{equation}
\begin{equation}\label{Bddf}
{\cal B}^{i}_{n}={1\over 2\alpha'}\oint {dz \over 2\pi i}\, \left(\,\p^{i}(q{\cdot}i\pa X)^{1/2} + {\p^{i}\over2} q{\cdot}\p \,q{\cdot}\pa \p \,(q{\cdot}i\pa X)^{-3/2}   - i\pa X^{i} \,q{\cdot}\p \, (q{\cdot}i\pa X)^{-1/2} \, \right) e^{inq{\cdot}X} 
\end{equation}
where $\p^i$ and $\pa X^i$ are the standard conformal primary operators of the world-sheet theory in the light-cone frame and $q^{\mu}$ is a light-like  ``reference''  momentum chosen to be of the form $(q^{-};q^{+}=0;\vec{q}=\vec{0})$.
These operators are well defined if the scalar product of the light-like momentum $q$ with the center of mass momentum of the string is proportional to the string tension, and in particular takes the value $1/(2\alpha')$. With this constraint they satisfy the Heisenberg Super-algebra of the oscillation modes
\begin{equation}\label{Halgebra}
[{\cal A}^{i}_{n},{\cal A}^{j}_{m}]=n\,\delta^{ij}\delta_{m+n,0}\quad;\quad \{{\cal B}^{i}_{n},{\cal B}^{j}_{m}\}=\delta^{ij}\delta_{m+n,0}
\end{equation}
that establishes a one to one correspondence with the oscillation modes ${\cal A}^i_n\sim \alpha^{i}_{n}$, ${\cal B}^i_n\sim b^{i}_{n}$. Acting with them on a specific vacuum state, one can generate the entire superstring physical spectrum whereby the operator action is reproduced by OPE techniques, as we will show in the next sub-section.

\subsection{OPE construction of general massive vertex operator}
To create states in the DDF formalism one needs to start from the tachyonic vacuum and produce states acting with creation operators. For each action there is an increase in momentum proportional to $q^\mu$ that automatically span the required mass level with the related combinations of transverse polarisations that match the relevant  degrees of freedom. The fact that one uses only transverse degrees of freedom makes this construction automatically BRST invariant but in principle not manifestly Lorentz covariant. The simplest vacuum choice to start with is a tachyonic vacuum in the canonical picture $q=-1$:
\begin{equation}\label{Vacuum}
\lim_{z\rightarrow 0 } \,e^{-\phi} e^{ip{\cdot}X}(z)\ket{0}=\ket{p}_{({-}1)}
\end{equation}
 where $p^2=1/(2\alpha')$ and ${\phi}$ is the boson for the superghost $(\beta,\gamma)$ system. A general state\footnote{In the rest of the paper we will tacitly assume the presence of the asymptotic limit in the creation of a state by the action of some operator.} is created by:
 \begin{equation}\label{masterState}
\left(\prod_{\ell} {\lambda_{n_{\ell}}\over n_{\ell} }{\cdot}{\cal A}_{-n_{\ell}}\right)  \left(\prod_{r} \nu_{m_{r}}{\cdot}{\cal B}_{-{m_{r}\over 2}}\right) \ket{p}_{(-1)}={\cal O}\left(   \{\lambda^{i}_{n_{\ell}}\},\{\nu^{i}_{m_{r}}\};\p^i,\pa X^i\right) \ket{k}_{(-1)}
\end{equation}
in fact there is a factorization between the total momentum operator that creates the momentum state $\ket{k}_{(-1)}$ and the remaining structure expressed in terms of a combination of primary operators. Where we would like to stress that the polarisation $\lambda$ is expressed in terms of c-numbers, the polarisation $\nu$ in terms of a-numbers. Our goal here is to find a compact expression for this combination and then write explicitly the general vertex operator. Before looking at the combination of primary operators, we want to show how to identify the mass level. This information is coded in the final momentum $k$  
\begin{equation}\label{totalMom}
k^\mu=p^{\mu}-\left(\sum_{\ell} n_{\ell} + \sum_{r} {m_{r}\over 2}\right)q^{\mu}
\end{equation}
using the constraint $p{\cdot}q=1/(2\alpha')$ one reproduces exactly the superstring mass-shell (\ref{m1}) and in order to restrict to the GSO-projected spectrum one has to replace $m$ with $2m-1$ imposing $m>0$ and to restrict the number of ${\cal B}$ operators to be odd giving rise to:
\begin{equation}\label{Mddf}
\mathbb{M}^2=2\left[\sum_{\ell} n_{\ell} + \sum_{\text{odd}\,r} (m_{r} - {1\over 2})\right]p{\cdot}q - p^2= {1\over \alpha'}\left[ \sum_{\ell}n_{\ell}+\sum_{\text{odd}\,r} (m_r {-}{1\over 2})-{1\over 2}\right]
\end{equation}

 For what concerns the operatorial structure ${\cal O}\left(   \{\lambda^{i}_{n_{\ell}}\},\{\nu^{i}_{m_{r}}\};\p^i,\pa X^i\right)$ of (\ref{masterState}), our strategy in deriving the closed form expression is to classify all the possible operators that come from the iterative action of ${\cal A}$ and ${\cal B}$ on the vacuum $\ket{p}_{(-1)}$. In particular it is possible to find a closed form expression because one can generate a finite number of different operators, concluding that this object admits a compact and elegant representation. To be more precise, we will obtain the operatorial structure of a general mass state as a particular case, or a specific single term, of a coherent vertex operator of states. In this way one is able not only to write a general vertex operator of a single state but also to include automatically, for each mass level, all the contributions with the relative degeneration.
 
At this point one has to compute the normal ordering that involves three types of operatorial actions: the ${\cal A}$ actions, ${\cal B}$ actions and mixed actions. The first one is
\begin{equation}\label{opeA}
{\lambda_n\over n}{\cdot}A_{-n}\, e^{ip{\cdot}X(z)}\,=\, {\lambda_n\over n}{\cdot}\left(A^{\pa X}_{-n} + A^{\p}_{-n}\right) \, e^{ip{\cdot}X(z)}
\end{equation}
where we have omitted the picture dependence of the vacuum and decomposed the ${\cal A}$ operator in two parts as follows
 \begin{equation}\label{Aoperator}
 \begin{split}
 &\lambda_{n}{\cdot}{\cal A}^{\pa X}_{-n}= \oint {dz\over 2\pi i}\, \lambda_{n}{\cdot}\pa X \,e^{-inq{\cdot}X}\,;\quad\lambda_{n}{\cdot}{\cal A}^{\p}_{-n}= \oint {dz\over 2\pi i}\, n\,q{\cdot}\p \lambda_{n}{\cdot}\p\, e^{-inq{\cdot}X}
 \end{split}
 \end{equation}
for simplicity we set $\alpha'=1/2$ here and in rest of the paper. Now one can proceed to the computation of the OPE's and in particular, just to set the notation, we compute a representative one and list below in Tab.(1)-Tab.(2) all the possible different operators that can be produced after the OPE. The representative OPE is the following

\begin{equation}\label{A1ope1}
:\oint {dz \over 2\pi i} \, {\lambda_{n}\over n}{\cdot}\pa X \,e^{-inq{\cdot}X}(z): :e^{ip{\cdot}X}(w):
\end{equation}

there are two contractions due to $\pa X$ and $e^{-inq{\cdot}X}$ with $e^{ip{\cdot}X}$ and the respective normal ordering parts that yield

\begin{equation}\label{A1ope2}
\oint {dz \over 2\pi i} \,(z{-}w)^{-n \,q{\cdot}p} \left[{\lambda_n{\cdot}p \over n (z{-}w)} + {\lambda_n \over n}{\cdot}\pa X(z)|_{z\rightarrow w}  \right] e^{-inq{\cdot}X(z)}|_{z\rightarrow w} \,e^{ip{\cdot}X}(w)
\end{equation}
using $p{\cdot}q=1$ and the Taylor expansion one gets
\begin{equation}\label{A1ope22}
\oint {dz \over 2\pi i} \, \left[{\lambda_n{\cdot}p \over n (z{-}w)^{n+1}} + \sum_{h=1}^{\infty}{\lambda_n \over n}{\cdot}{\pa^{h} X(w) \over (h-1)!}(z{-}w)^{n{+}h{-}1}  \right] \exp\left(-in\sum_{\ell=1}^{\infty} {q{\cdot}\pa^{\ell}X \over \ell (\ell{-}1)!} \right) \,e^{i(p-nq){\cdot}X}(w)
\end{equation}
and computing the integral, the final expression becomes 
\begin{equation}
{\lambda_n \over n}{\cdot}\widetilde{{\cal P}_{n}} \,e^{i(p-nq){\cdot}X}(w)=\left({\lambda_n \over n}{\cdot}p \, {\cal Z}_{n}\left[ {\cal U}^{(n)}_{s} \right] {+} \sum_{h=1}^{n} {\lambda_n \over n}{\cdot}{\pa^{h} X(w) \over (h-1)!} \, {\cal Z}_{n{-}h}\left[ {\cal U}^{(n)}_{s} \right]  \right) \, e^{i(p-nq){\cdot}X}(w)
\label{Phat}
\end{equation}

where $\widetilde{{\cal P}^i_{n}}= p^i {\cal Z}_n + {\cal P}^i_{n}$ with ${\cal Z}_{n}$ is a cycle index polynomial of the symmetric group with operatorial argument 
\begin{equation}\label{Zargument}
{\cal U}^{(n)}_s= -in{q{\cdot}\pa^s X\over (s-1)! }
\end{equation}
The properties of the ${\cal Z}_n$ polynomials are summarised in Appendix A.
Relying on similar considerations one can compute the remaining OPE's whose explicit operatorial structures are listed in Appendix B.
\begin{table}[!ht]
\center
\begin{tabular}{ | c || c |}
  ${\lambda_n \over n}{\cdot}{\cal A}^{\pa X}_{-n}$ & \\ \hline  & \\ 
  ${\lambda_{n}\over n}{\cdot}\widetilde{{\cal P}}_{n} \,e^{i (p{-}nq){\cdot}X}$ & $e^{i p{\cdot}X}$ \\\hline & \\ 
  $\left( {\lambda_{n} {\cdot}\lambda_m \over n m}{\cal S}_{m,n} +{\lambda_{m}\over m}{\cdot}\widetilde{{\cal P}}_{m}  {\lambda_{n}\over n}{\cdot}\widetilde{{\cal P}}_{n}\right) e ^{i [p{-}(n{+}m)q]{\cdot}X}$ & ${\lambda_{m}\over m}{\cdot}\widetilde{{\cal P}}_{m} e ^{i (p{-}mq){\cdot}X}$ \\ & \\ 
  ${\lambda_{n}\over n}{\cdot}\widetilde{{\cal P}}_{n} \,\lambda_{m}{\cdot}{ {\cal E}}_{m-1}\,e ^{i [p{-}(n{+}m)q]{\cdot}X}$ & $ \lambda_{m}{\cdot}{ {\cal E}}_{m-1} e ^{i (p{-}mq){\cdot}X}$\\
 \toprule
\end{tabular}
\caption{ first type of ${\cal A}$ actions useful to generate all the different operators that appear in the construction. }
\label{tabADX}
\end{table}
\begin{table}[!ht]
\center
\begin{tabular}{ | c || c |}
  ${\lambda_n \over n}{\cdot}{\cal A}^{\p}_{-n}$ &   \\ \hline  & \\ 
 $ \lambda_{n}{\cdot}{ {\cal E}}_{n-1} e ^{i (p{-}nq){\cdot}X}$ & $e^{i p{\cdot}X}$ \\\hline & \\ 
 ${\lambda_{m}\over m}{\cdot}\widetilde{{\cal P}}_{m} \,\lambda_{n}{\cdot}{ {\cal E}}_{n-1}\,e ^{i [p{-}(n{+}m)q]{\cdot}X}$ & ${\lambda_{m}\over m}{\cdot}\widetilde{{\cal P}}_{m} e ^{i (p{-}mq){\cdot}X}$ \\ & \\ 
  $\left( {\lambda_{n}{\cdot}\lambda_{m}} {\cal F}_{m{-}1,n} + \lambda_n{\cdot}{\cal E}_{n{-}1}\lambda_m{\cdot}{\cal E}_{m{-}1}  \right) e ^{i [p{-}(n{+}m)q]{\cdot}X}$ & $ \lambda_{m}{\cdot}{ {\cal E}}_{m-1} e ^{i (p{-}mq){\cdot}X}$\\
 \toprule
\end{tabular}
\caption{ last type of ${\cal A}$ actions useful to generate all the different operators that appear in the construction. }
\label{tabAY}
\end{table}
In order to investigate the action of the ${\cal B}$ operators one can proceed as above and see how many different operators can appear. The first OPE to compute is
\begin{equation}\label{Boperator}
\nu_{n}{\cdot}{\cal B}_{-n{+}{1\over 2}}\,e^{ip{\cdot}X}= \nu_{n}{\cdot}\left({\cal B}^{(1/2)}_{-n{+}{1\over 2}}+{\cal B}^{(-3/2)}_{-n{+}{1\over 2}}+{\cal B}^{(-1/2)}_{-n{+}{1\over 2}}\right)\,e^{ip{\cdot}X}
\end{equation}
where we have once again separated three different contributions generated by:
\begin{eqnarray}\label{Boperators}
&&\nu_{n}{\cdot}{\cal B}^{(1/2)}_{-n{+}{1\over 2}}=\oint {d z \over 2\pi i} \,\nu_{n}{\cdot}\p \,(iq{\cdot}\pa X)^{1/2}\, e^{-i(n{-}1/2)q{\cdot}X}\\
&&\nu_{n}{\cdot}{\cal B}^{(-3/2)}_{-n{+}{1\over 2}}=\oint {d z \over 2\pi i} \,{\nu_{n}\over 2}{\cdot}\p \,q{\cdot}\p\,q{\cdot}\pa \p  \,(iq{\cdot}\pa X)^{-3/2} \,e^{-i(n{-}1/2)q{\cdot}X}\\
&&\nu_{n}{\cdot}{\cal B}^{(-1/2)}_{-n{+}{1\over 2}}=-\oint {d z \over 2\pi i} \,\nu_{n}{\cdot}i\pa X \,q{\cdot}\p \,(iq{\cdot}\pa X)^{-1/2} \,e^{-i(n{-}1/2)q{\cdot}X}
\end{eqnarray}

The representative OPE that we explicitly compute is the following
\begin{eqnarray}\label{B1ope}
&&\nu_{n}{\cdot}{\cal B}^{(1/2)}_{-n{+}1/2}\,e^{ip{\cdot}X}(w)=:\oint {d z \over 2\pi i} \,\nu_{n}{\cdot}\p \,(iq{\cdot}\pa X)^{1/2}\, e^{-i(n{-}1/2)q{\cdot}X}(z):\,:e^{ip{\cdot}X}(w): 
\end{eqnarray}
and all the others OPE's are listed in Tab(3)-Tab(5). 
The first comment is that one can realise how important is the GSO projection in this formalism. In fact looking at the OPE, the very choice of the GSO projection used in (\ref{Mddf}) exclude automatically the possibility of branch cuts.
Another comment is that there is a non standard OPE to perform which involves an operator of the form $(iq{\cdot}\pa X)^{1/2}$. One possible way to compute it, is to use the Schwinger parametrization:
\begin{equation}\label{SchPG12}
(iq{\cdot}\pa X)^{1/2}(z)= {1\over \Gamma(-{1\over 2})}\int_{0}^{\infty}d\eta\,\eta^{-1/2{-}1} e^{-\eta\, iq{\cdot}\pa X(z)}
\end{equation}
Then the OPE (\ref{B1ope}) yields
\begin{equation}\label{B1ope2}
\oint {dz \over 2\pi i} \, \nu_{n}{\cdot}\p(z)|_{z\rightarrow w} \Big[(z{-}w)^{-1} {+} iq{\cdot}\pa X(z)|_{z\rightarrow w}  \Big]^{1/2} (z{-}w)^{-n{+}1/2} \,e^{-i(n{-}1/2)q{\cdot}X(z)}|_{z\rightarrow w} \,e^{ip{\cdot}X}(w)
\end{equation}
and using Taylor expansion the expression becomes
\begin{equation}\label{B1ope3}
\oint {dz \over 2\pi i} \, \sum_{h=1}^{\infty}{\nu_{n}\over h!}{\cdot}{\pa^{h} \p (w)} \left(1{+}\sum_{\ell=1}^{\infty} {\cal U}^{(-1)}_{\ell}(z{-}w)^{\ell}\right)^{1/2} (z{-}w)^{-n{+}h}\, \exp\left(\sum_{v=1}^{\infty}{{\cal U}^{(n{-}1/2)}_v \over v(z{-}w)^{-v}} \right) e^{i[p{-}(n{-}1/2)q]{\cdot}X}
\end{equation}
where we have used the same notation as in (\ref{Zargument}). The evaluation of the integral gives a new polynomial structure $\cal Q$ that is shown and explained in Appendix A. The final result is:
\begin{equation}\label{B1ope1Ris}
\nu_{n}{\cdot}{\cal O}^{\p}_{n{-}1}e^{i[p{-}(n{-}1/2)q]{\cdot}X}(w)=\sum_{h=0}^{n{-}1} {\nu_{n}\over h!}{\cdot}\pa^{h} \p \, {\cal Q}_{n{-}1{-}h}^{(1/2,n)}\,e^{i[p{-}(n{-}1/2)q]{\cdot}X}(w)
\end{equation}

In order to compute the remaining OPE one can adopt the same strategy and adapt to the specific cases the Schwinger parametrization:
\begin{equation}\label{SChW para}
{\cal O}^{k}={1\over \Gamma(-k)}\int_{0}^{\infty}d\eta\, \eta^{-k-1} e^{-\eta\,{\cal O}}
\end{equation}

\begin{table}[!ht]
\center
\begin{tabular}{| c || c |}
  $\nu_n{\cdot}{\cal B}^{(1/2)}_{-n+{1\over 2}}$ &     \\ \hline & \\ 
 $ \nu_n{\cdot}{\cal O}^{\p}_{n-1} e^{i [p{-}(n{-}1/2)q]{\cdot}X} $ & $e^{i p{\cdot}X} $ \\\hline & \\ 
 $\left( \nu_m{\cdot}\nu_n \, {\cal K}^{({1\over 2}\,,{1\over 2})}_{m{-}1,n} + \nu_n{\cdot}{\cal O}^{\p}_{n{-}1} \nu_m{\cdot}{\cal O}^{\p}_{m{-}1}   \right) e^{i [p{-}(n{+}m{-}1)q]{\cdot}X} $ &  $ \nu_m{\cdot}{\cal O}^{\p}_{m{-}1} e^{i [p{-}(m{-}1/2)q]{\cdot}X} $ \\ 
 & \\ 
 $\left( \nu_m{\cdot}\nu_n \, {\cal K}^{(-{3\over 2}\,,{1\over 2})}_{m{-}2,n{-}1} + \nu_n{\cdot}{\cal O}^{\p}_{n{-}1} \nu_m{\cdot}{\cal O}^{\p\p\p}_{m{-}2}   \right)  e^{i [p{-}(n{+}m{-}1)q]{\cdot}X} $ & $\nu_m{\cdot}{\cal O}^{\p\p\p}_{m-2} e^{i [p{-}(m{-}1/2)q]{\cdot}X}$\\
  & \\ 
 $-\nu_m{\cdot}{\cal O}^{\p p}_{m-1} \nu_n{\cdot}{\cal O}^{\p}_{n-1} e^{i [p{-}(n{+}m{-}1)q]{\cdot}X}$ & $-\nu_m{\cdot}{\cal O}^{\p p}_{m-1} e^{i [p{-}(m{-}1/2)q]{\cdot}X} $\\
  & \\ 
 $-\nu_m{\cdot}{\cal O}^{\p \pa X}_{m-1} \nu_n{\cdot}{\cal O}^{\p}_{n-1}e^{i [p{-}(n{+}m{-}1)q]{\cdot}X}$ & $-\nu_m{\cdot}{\cal O}^{\p \pa X}_{m-1} e^{i [p{-}(m{-}1/2)q]{\cdot}X} $\\
 \toprule
\end{tabular}
\caption{ actions of the first type  of ${\cal B}$ useful to generate all the possible different operators }
\label{tabB1/2}
\end{table}
\begin{table}[!ht]
\center
\begin{tabular}{| c || c |}
 $\nu_n{\cdot}{\cal B}^{(-3/2)}_{-n+{1\over 2}}$ &     \\ \hline & \\ 
$\nu_n{\cdot}{\cal O}^{\p\p\p}_{n-1} e^{i [p{-}(n{-}1/2)q]{\cdot}X}$  & $e^{i p{\cdot}X} $ \\\hline & \\ 
 $\left(\nu_m{\cdot}\nu_n\,  {\cal K}^{({1\over 2}\,,-{3\over 2})}_{m{-}1,n{-}1} + \nu_n{\cdot}{\cal O}^{\p\p\p}_{n{-}2} \nu_m{\cdot}{\cal O}^{\p}_{m{-}1} \right)  e^{i [p{-}(n{+}m{-}1)q]{\cdot}X}$  & $ \nu_m{\cdot}{\cal O}^{\p}_{m{-}1} e^{i [p{-}(m{-}1/2)q]{\cdot}X} $ \\ 
 & \\ 
 $\left( \nu_m{\cdot}\nu_n\,  {\cal K}^{(-{3\over 2}\,,-{3\over 2})}_{m{-}2,n} + \nu_n{\cdot}{\cal O}^{\p\p\p}_{n{-}2}\nu_m{\cdot}{\cal O}^{\p\p\p}_{m{-}2} \right)  e^{i [p{-}(n{+}m{-}1)q]{\cdot}X}$  & $\nu_m{\cdot}{\cal O}^{\p\p\p}_{m-2} e^{i [p{-}(m{-}1/2)q]{\cdot}X}$\\
  & \\ 
 $-\nu_m{\cdot}{\cal O}^{\p p}_{m-1} \nu_n{\cdot}{\cal O}^{\p\p\p}_{n-1} e^{i [p{-}(n{+}m{-}1)q]{\cdot}X}$  & $-\nu_m{\cdot}{\cal O}^{\p p}_{m-1} e^{i [p{-}(m{-}1/2)q]{\cdot}X} $\\
  & \\ 
$-\nu_m{\cdot}{\cal O}^{\p \pa X}_{m-1} \nu_n{\cdot}{\cal O}^{\p\p\p}_{n-1} e^{i [p{-}(n{+}m{-}1)q]{\cdot}X}$  & $-\nu_m{\cdot}{\cal O}^{\p \pa X}_{m-1} e^{i [p{-}(m{-}1/2)q]{\cdot}X} $\\
 \toprule
\end{tabular}
\caption{ actions of the second type of ${\cal B}$ useful to generate all the possible different operators}
\label{tabB-3/2}
\end{table}
\begin{table}[!ht]
\center
\begin{tabular}{| c || c |}
  $\nu_n{\cdot}{\cal B}^{(-1/2)}_{-n+{1\over 2}}$ &   \\ \hline & \\ 
 $-\nu_n{\cdot}\left({\cal O}^{\p p}_{n-1}{+}{\cal O}^{\p \pa X}_{n-1} \right) e^{i [p{-}(n{-}1/2)q]{\cdot}X}$ & $e^{i p{\cdot}X} $ \\\hline & \\ 
 $-\nu_n{\cdot}\left({\cal O}^{\p p}_{n{-}1}{+}{\cal O}^{\p \pa X}_{n{-}1} \right)  \nu_m{\cdot}{\cal O}^{\p}_{m{-}1} e^{i [p{-}(m{+}n{-}1)q]{\cdot}X}$ & $ \nu_m{\cdot}{\cal O}^{\p}_{m{-}1} e^{i [p{-}(m{-}1/2)q]{\cdot}X} $ \\ 
 & \\ 
 $-\nu_n{\cdot}\left({\cal O}^{\p p}_{n-1}{+}{\cal O}^{\p \pa X}_{n-1} \right)  \nu_m{\cdot}{\cal O}^{\p\p\p}_{m-2} e^{i [p{-}(n{+}m{-}1)q]{\cdot}X}$ & $\nu_m{\cdot}{\cal O}^{\p\p\p}_{m-2} e^{i [p{-}(m{-}1/2)q]{\cdot}X}$\\
  & \\ 
 $\nu_n{\cdot}\left({\cal O}^{\p p}_{n-1}{+}{\cal O}^{\p \pa X}_{n-1} \right)  \nu_m{\cdot}{\cal O}^{\p p}_{m-1} e^{i [p{-}(n{+}m{-}1)q]{\cdot}X}$ & $-\nu_m{\cdot}{\cal O}^{\p p}_{m-1} e^{i [p{-}(m{-}1/2)q]{\cdot}X} $\\
  & \\ 
 $\left\{\nu_n{\cdot}\left({\cal O}^{\p p}_{n-1}{+}{\cal O}^{\p \pa X}_{n-1} \right)  \nu_m{\cdot}{\cal O}^{\p \pa X}_{m-1} {+}\nu_m{\cdot}\nu_n \, {\cal K}^{(-{1\over 2}\,,-{1\over 2})}_{m{-}1,n} \right\} e^{i [p{-}(n{+}m{-}1)q]{\cdot}X}$ & $-\nu_m{\cdot}{\cal O}^{\p \pa X}_{m-1} e^{i [p{-}(m{-}1/2)q]{\cdot}X} $\\
 \toprule
\end{tabular}
\caption{  actions of the last type of ${\cal B}$ useful to generate all the possible different operators }
\label{tabB-1/2}
\end{table}

Finally the last new operators that appear are the results of the combined action of ${\cal A}$ and ${\cal B}$ as listed in the last three tables Tab.(6)-Tab(8).
\begin{table}[!ht]
\center
\begin{tabular}{| c || c |}
 $\nu_n{\cdot}{\cal B}^{(1/2)}_{-n+{1\over 2}}$ &    \\ \hline & \\ 
 $\nu_n{\cdot}{\cal O}^{\p}_{n-1}\,{\lambda\over m}{\cdot}\widetilde{{\cal P}}_m  e^{i [p{-}(n{+}m{-}1/2)]{\cdot}X} $ & $ {\lambda_m\over m}{\cdot}\widetilde{{\cal P}}_m e^{i (p{-}mq){\cdot}X} $ \\\hline & \\ 
 $\left(\lambda_n{\cdot}{{\cal E}}_{n{-}1}\, \nu_{m}{\cdot}{\cal O}^{\p}_{m{-1}} {+} {\lambda_n {\cdot}\nu_m}{\cal Y}^{({1\over 2})}_{n{-}1,m} \right) e^{i [p{-}(n{+}m{-}1/2)]{\cdot}X}$ & ${\zeta_m}{\cdot}{ {\cal E}}_{m-1} e^{i (p{-}mq){\cdot}X}$ \\ 
 \toprule
\end{tabular}
\caption{ first type of combined actions of ${\cal A}$ and ${\cal B}$ }
\label{tabB1/2_2}
\end{table}
\begin{table}[!ht]
\center
\begin{tabular}{| c || c |}
 $\nu_n{\cdot}{\cal B}^{(-3/2)}_{-n+{1\over 2}}$ &    \\ \hline &\\ 
 $\nu_n{\cdot}{\cal O}^{\p\p\p}_{n-1}\,{\lambda_m\over m}{\cdot}\widetilde{{\cal P}}_m  e^{i [p{-}(n{+}m{-}1/2)]{\cdot}X} $   & $ {\lambda_m\over m}{\cdot}\widetilde{{\cal P}}_m e^{i (p{-}mq){\cdot}X} $ \\\hline & \\ 
 $\left(\lambda_n{\cdot}{{\cal E}}_{n{-}1} \nu_{m}{\cdot}{\cal O}^{\p\p\p}_{m{-}2} {+} {\lambda_n {\cdot}\nu_m}{\cal Y}^{(-{3\over 2})}_{n{-}1,m{-}1} \right) e^{i [p{-}(n{+}m{-}1/2)]{\cdot}X}$ & ${\zeta_m}{\cdot}{ {\cal E}}_{m{-}1} e^{i (p{-}mq){\cdot}X}$ \\ 
 \toprule
\end{tabular}
\caption{ second type of combined actions of ${\cal A}$ and ${\cal B}$ }
\label{tabB-3/2_2}
\end{table} 
\begin{table}[!ht]
\center
\begin{tabular}{| c || c |}
 $\nu_n{\cdot}{\cal B}^{(-1/2)}_{-n+{1\over 2}}$ &    \\ \hline & \\ 
 $-\left\{\nu_n{\cdot}\left({\cal O}_{n{-}1}^{\p p}{+}{\cal O}_{n{-}1}^{\p \pa X}\right){\lambda_m\over m}{\cdot}\widetilde{{\cal P}}_m {+} {1\over m}{\lambda_{m}{\cdot}\nu_n}{\cal Y}^{(-{1\over 2})}_{n,m{-}1} \right\}  e^{i [p{-}(n{+}m{-}1/2)]{\cdot}X}$  & $ {\lambda_m\over m}{\cdot}\widetilde{{\cal P}}_m e^{i (p{-}mq){\cdot}X} $ \\\hline & \\ 
$-\nu_n{\cdot}\left({\cal O}_{n{-}1}^{\p p}{+}{\cal O}_{n{-}1}^{\p \pa X}\right){\lambda_m}{\cdot}{ {\cal E} }_{m-1} e^{i [p{-}(n{+}m{-}1/2)]{\cdot}X} $ & ${\lambda_m}{\cdot}{ {\cal E}}_{m-1} e^{i (p{-}mq){\cdot}X}$ \\ 
 \toprule
\end{tabular}
\caption{ last type of combined actions of ${\cal A}$ and ${\cal B}$ }
\label{tabB-1/2_2}
\end{table}
As we already mentioned the vertex operators found in this formalism are automatically BRST invariant but not manifestly Lorentz covariant. After the OPE construction one can show that there exists a ``covariantization'' map that is represented by the following rules
\begin{equation}\label{Cov map}
\lambda^{i}_{n}\rightarrow \zeta_{n}^{\mu}= \lambda_{n}^{i}\left(\delta^{i \mu} - p^i q^\mu \right)\,; \quad \nu^{i}_{n}\rightarrow {\Upsilon}_{n}^{\mu}= \nu_{n}^{i}\left(\delta^{i \mu} - p^i q^\mu \right)
\end{equation}
using this map, general structures as in (\ref{masterState}) assume a manifestly covariant representation
\begin{equation}\label{Cov struct}
{\cal O}\left(   \{\lambda^{i}_{n_{\ell}}\},\{\nu^{i}_{m_{r}}\};\p^i,\pa X^i\right) \rightarrow {\cal O}\left(   \{\zeta^{\mu}_{n_{\ell}}\},\{{\Upsilon}^{\mu}_{m_{r}}\};\p^\mu,\pa X^\mu\right) 
\end{equation}

To conclude this subsection we give the general formula of the normal ordered version of a NS vertex operator of generic mass in the canonical superghost picture. With the help of the OPE rules derived before, one can write down the closed form for the action of a generic number of the two types of DDF operators on the tachyonic vertex operator, in relation with (\ref{masterState}), that takes the following form

\begin{equation}
\begin{split}
&{\cal A}^{i_1}_{n_{1}}...{\cal A}^{i_{g1}}_{n_{g1}}{\cal B}^{j_1}_{m_{1}}...{\cal B}^{j_{g2}}_{m_{g2}} \, \,e^{-\phi} e^{ip{\cdot}X}(z)=\\
&(-)^{g_2}e^{-\phi}\sum_{a_1=0}^{[g_{1}/2]}  \sum_{a_2=0}^{[g_{2}/2]} \sum_{\pi_{1}\in \mathbf{G}_{1}}  \sum_{\pi_{2}\in \mathbf{G}_{2}}\prod_{h_{1}=1}^{a_{1}}\prod_{h_{2}=1}^{a_{2}}    \delta^{i_{\pi_{1}(h_1)} j_{\pi_{2}(h_2)} } f^{(\lambda \nu)}_{n_{\pi_{1}(h_{1})} m_{\pi_{2}(h_{2})} }     \hspace{-1mm}    \prod_{\ell_{1}=1}^{a_{1}} \delta^{i_{\pi_{1}(2\ell_{1}{-}1)} i_{\pi_{1}(2\ell_{1})} } f^{(\lambda \lambda)}_{n_{\pi_{1}(2\ell_{1}{-}1)} n_{\pi_{1}(2\ell_{1})} }\\
&\prod_{s_{1}=2a_{1}{+}1}^{g_{1}} \hspace{-4mm} f^{(\lambda)i_{\pi_{1}(s_{1})}}_{n_{\pi_{1}(s_{1})}}  \prod_{\ell_{2}=1}^{a_{2}} \delta^{j_{\pi_{2}(2\ell_{2}{-}1)} j_{\pi_{2}(2\ell_{2})} } f^{(\nu \nu)}_{m_{\pi_{2}(2\ell_{2}{-}1)} m_{\pi_{2}(2\ell_{2})} }  \hspace{-3mm} \prod_{s_{2}=2a_{2}{+}1}^{g_{2}} f^{(\lambda)j_{\pi_{2}(s_{2})}}_{m_{\pi_{2}(s_{2})}}\,e^{i\big[p{-}\left(\sum_{r=1}^{g_{1}}n_{r} {+} \sum_{v=1}^{g_{2}}m_{v}  \right)q\big]{\cdot}X}(z)
\end{split}
\label{eq:GMS}
\end{equation}
where $\mathbf{G}_{k}=\mathbf{S}ymm_{k}/\mathbb{Z}_k$ is the symmetric group of k-elements quotient with $\mathbb{Z}_k$ and the $f^{()}$ operators are
\begin{equation}
\begin{split}
  &f^{(\lambda \nu)}_{n m}={\cal Y}^{({1\over 2})}_{n{-}1,m}+{\cal Y}^{(-{1\over 2})}_{n,m{-}1}+{\cal Y}^{(-{3\over 2})}_{n{-}1,m{-}1}\,; \quad\quad\quad f^{(\lambda \lambda)}_{n_1 n_2}= {\cal S}_{n_{1},n_{2}} + {\cal F}_{n_{1}{-}1,n_{2}}\\
  & f^{(\nu)j}_{m}={\cal O}^{i,\p}_{m{-}1}+{\cal O}^{i,\p\p\p}_{m{-}2}+{\cal O}^{i,\p p}_{m{-}1}+{\cal O}^{i,\p \pa X}_{m{-}1}\,;\quad f^{(\lambda)i}_{n}= {\cal P}^{i}_{n}+{\cal E}^{i}_{n{-}1}\\
  &f^{(\nu \nu)}_{m_1 m_2}= {\cal K}^{(-{1\over 2}),(-{1\over 2})}_{m_{1}{-}1,m_{2}{-}1}+{\cal K}^{(-{3\over 2}),({1\over 2})}_{m_{1}{-}2,m_{2}}+{\cal K}^{({1\over 2}),(-{3\over 2})}_{m_{1}{-}1,m_{2}{-}1}+ {\cal K}^{(-{3\over 2}),(-{3\over 2})}_{m_{1}{-}2,m_{2}}+ {\cal K}^{({1\over 2}),({1\over 2})}_{m_{1}{-}1,m_{2}} 
  \end{split}
  \label{eq:GMS_lab}
  \end{equation}  
where the explicit structures are listed in Appendix B.

\subsection{Definition of NS-coherent vertex operator in canonical picture}
A NS coherent state can be created as a superposition of the Grassmann even and Grassmann odd creation operators acting on a certain vacuum state, in particular the notion of coherence is related to the fact that the state will be an eigenstate of the annihilation operators. In this context there are two different kinds of annihilation operators and one can  construct a state that is coherent under the action of both annihilation operators.
Having in mind the OPE rules derived in the previous subsection and the map (\ref{Cov map}) one can define a NS coherent vertex operator in the canonical picture, automatically BRST invariant and Lorentz covariant, simply by:

\begin{equation}\label{defCoheVert}
{\cal V}^{(NS)}_{{\cal C}\,\text{full}\,(-1)}(z)=\exp\left(\sum_{n=1}^{\infty} {\zeta_n \over n}{\cdot} {\cal A}_{-n}\right) \, \rho^{(NS)+}_{\text{GSO}}\,\exp\left(\sum_{m=1}^{\infty} {\Upsilon_{m}}{\cdot}{\cal B}_{-m+{1\over 2}}\right) \, e^{-\phi}\,e^{ip{\cdot}X}(z)  
\end{equation}
where the GSO projectors are
\begin{equation}\label{GSO proj}
\rho^{(NS)+}_{GSO}={1+(-)^{G}\over 2}\,;\quad \rho^{(NS)-}_{GSO}={1-(-)^{G}\over 2}\,;\quad G=\sum {\cal B}_{-m{+}1/2} {\cal B}_{m{-}1/2}+1
\end{equation}
the operator $(-)^G$ has $-1$ eigenvalue on the tachyon state $e^{- \phi}\,e^{ip{\cdot}X}(z)$, that we have chosen in the canonical superghost picture. The definition (\ref{defCoheVert}) can be made more transparent, in the sense that one can represent it in terms of the primary conformal fields that appear in the DDF operators performing the normal ordering between the DDF operators and the tachyonic state. We stress that one goal is the computation of the normal ordered version of (\ref{defCoheVert}) that gives rise to a coherent vertex operator in which instead of the dependence on the creation operators there are operatorial structures  that level by level reproduce the expected BRST invariant and manifestly covariant vertex operators. 
The identification of the vertex operator with the corresponding state is given by
\begin{equation}\label{Cohe state}
\ket{{\cal{C}}(\{\zeta\},\{\Upsilon\})}=\exp\left(\sum_{n=1}^{\infty} {\zeta_n \over n}{\cdot} {\cal A}_{-n}\right) \, \rho^{(NS)+}_{\text{GSO}}\,\exp\left( \sum_{m=1}^{\infty} {\Upsilon_{m}}{\cdot}{\cal B}_{-m+{1\over 2}}\right) \ket{p}_{(-1)}
\end{equation}
this state is coherent under the action of a generic lowering operator ${\cal{A}}_\ell^{k}$ :
\begin{equation}\label{coherence}
{\cal{A}}_\ell^{k} \ket{{\cal{C}}(\{\zeta\},\{\Upsilon\})}=\zeta_{\ell}^{k}\ket{{\cal{C}}(\{\zeta\},\{\Upsilon\})}
 \end{equation} 
and is a squeezed coherent state with respect to the ${\cal{B}}$ dependence:
\begin{equation}\label{sqeezed}
{\cal{B}}_{\ell}^{i}{\cal{B}}_{v}^{j} \ket{{\cal{C}}(\{\zeta\},\{\Upsilon\})}=\Upsilon^i_{\ell} \Upsilon^j_v  \ket{{\cal{C}}(\{\zeta\},\{\Upsilon\})}
\end{equation}
The state (\ref{Cohe state}) can be normalized multiplying the factor
\begin{equation}\label{normalization}
 \exp\left(-\sum_{n=1}^{\infty} \zeta_{n}^{*}{\cdot}\zeta_{n} - \sum_{m=1}^{\infty} \Upsilon_{m}^{*}{\cdot}\Upsilon_{m}\right)
 \end{equation} 
and in order to obtain a well defined state one has to impose the convergence constraints
\begin{equation}
\sum_{n=1}^{\infty} \zeta_{n}^{*}{\cdot}\zeta_{n} < \infty \,\quad \sum_{m=1}^{\infty} \Upsilon_{m}^{*}{\cdot}\Upsilon_{m}<\infty
\label{eq:finite_norm}
\end{equation}
To construct explicitly the full vertex operator one can study, separately the action of the ${\cal A}$ and ${\cal B}$ operators, and after that the combined action. This is the way in which we proceed.
The exponential part involving only ${\cal A}$ operators yields
\begin{equation}\label{AcoheVert}
\exp\left(\sum_{n=1}^{\infty} {\zeta_n\over n}{\cdot}{\cal A}_{-n}\right) \,e^{-\phi}e^{ip{\cdot}X}=e^{{\cal O}_{1} + {\cal O}_{2}+{\cal O}_{3}+{\cal O}_{4}} \,e^{-\phi}e^{ip{\cdot} X}
\end{equation}
where the operators are :
\begin{eqnarray}\label{AoperatorsCoher}
&&{\cal O}_1= \sum_{m,n=1} {\zeta_m {\cdot}\zeta_n \over 2 m n}\, {\cal S}_{m,n} \,e^{-i(m+n)q{\cdot}X} \,; \quad {\cal O}_2= \sum_{n=1} {\zeta_n \over n}{\cdot} {\cal P}_{n} \,e^{-inq{\cdot}X} \\
&&{\cal O}_3= \sum_{n=1} {\zeta_n}{\cdot}  { {\cal E}}_{n-1} \,e^{-inq{\cdot}X}\,; \quad \hspace*{1,7cm}{\cal O}_4= \sum_{m,n=1} {\zeta_m {\cdot}\zeta_n}\, {\cal F}_{n-1,m} \,e^{-i(m+n)q{\cdot}X} 
\end{eqnarray}

From the Grassmann odd operators one gets:
\begin{equation}\label{BcoherVert}
\rho^{(NS)+}_{\text{GSO}}\,\exp\left(\sum_{m=1}^{\infty} {\Upsilon_{m}}{\cdot}{\cal B}_{-m+{1\over 2}}\right) \,e^{-\phi}e^{ip{\cdot}X}=e^{{\cal H}} \sinh\left({\cal H}_1-{\cal H}_2 \right) e^{-\phi}e^{ip{\cdot}X}
\end{equation}
where
\begin{eqnarray}\label{BoperatorsCoher}
&&{\cal H}_{1}=\sum_{n=1}{\Upsilon}_{n}{\cdot}\left({\cal O}^{\p}_{n{-}1}+ {\cal O}^{\p\p\p}_{n{-}2} \right) \,e^{-i(n-1/2)q{\cdot}X}        \,; \quad {\cal H}_{2}=\sum_{n=1} {\Upsilon}_{n}{\cdot} {\cal O}^{\p \pa X}_{n{-}1} \,e^{-i(n-1/2)q{\cdot}X} \nonumber \\
&&{\cal H}=\sum_{m,n=1} {{\Upsilon}_{n}{\cdot}{\Upsilon}_{m}\over 2} \left( {\cal K}^{(-\frac{1}{2}),(-\frac{1}{2})}_{n{-}1,m{-}1}+{\cal K}^{(-\frac{3}{2}),(-\frac{3}{2})}_{n{-}2,m}+{\cal K}^{(-\frac{3}{2}),(\frac{1}{2})}_{n{-}2,m{-}1} + {\cal K}^{(\frac{1}{2}),(-\frac{3}{2})}_{n{-}1,m{-}1}+{\cal K}^{(\frac{1}{2}),(\frac{1}{2})}_{n{-}1,m}\right)\,e^{-i(m+n-1)q{\cdot}X} \nonumber \\
\end{eqnarray}
Finally from the combined action one gets an exponential term with argument:
\begin{equation}\label{com actions}
{\cal I}=\sum_{n,m=1} \zeta_{m}{\cdot}{\Upsilon}_n \left( {\cal Y}_{n{-}1,m}^{({1\over 2})} {+}{\cal Y}_{n,m{-}1}^{(-{3\over 2})}{-}{1\over m}{\cal Y}_{n{-}1,m}^{(-{1\over 2})}  \right) e^{-i(n{+}m{-}1/2)q{\cdot}X }
\end{equation}

The manifestly BRST invariant coherent vertex operator in the NS-sector, after the normal ordering, takes the form
\begin{equation}\label{NScohe}
\begin{split}
{\cal V}_{{\cal C}\,\text{full}\,(-1)}^{(NS)}(z)&=e^{-\phi}e^{{\cal O}_{1} {+} {\cal O}_{2}{+}{\cal O}_{3}{+}{\cal O}_{4}{+}{{\cal H}}{+}{\cal I}} \sinh\left( {\cal H}_1{-}{\cal H}_2 \right)   e^{ip{\cdot}X}
\end{split}
\end{equation}
The operators identified by the polarisation structures $\zeta{\cdot}\zeta$, ${\Upsilon}{\cdot}{\Upsilon}$ and $\zeta{\cdot}{\Upsilon}$ are new operators, and therefore new contributions, originating from the coherence effects due to the superposition of all states. The fact that these effects are quadratic in the polarizations reflects their massive nature. One can put these new effects to zero choosing complex polarisations and working in the frame in which there are only familiar string contributions using the following vertex operator:
\begin{equation}\label{simple NS coher}
\begin{split}
&{\cal V}_{{\cal C} (-1)}^{(NS)}(z)=e^{-\phi}e^{ {\cal O}_{2}{+}{\cal O}_{3}} \sinh\left( {\cal H}_1{-}{\cal H}_2 \right)   e^{ip{\cdot}X}(z)
\end{split}
\end{equation}
 In what follows, in order to simplify the notation without loss of generality, we will proceed the analysis with the last vertex operator.
 To make contact with the nature of the vertex operator it is useful look at the first mass levels and recognize the respective vertex operators. To do this one can start with the mass shell condition of the coherent state 
\begin{equation}
\begin{aligned}
\al \mathbb{M}^2& = N + M -\frac{1}{2} = N + M^{o,e} + M^{e,o} - \frac{1}{2}
\end{aligned}
\label{eq:mass2_q-1}
\end{equation}
and for simplicity take the following representation of the vertex operator
\begin{equation}
\begin{aligned}
{\cal V}^{(NS)}_{{\cal C}(-1)}(z) &= e^{-\phi} e^{{\cal O}_2 + {\cal O}_3}\Big(\sinh({\cal{H}}_1)\cosh({\cal{H}}_2) - \sinh({\cal{H}}_2)\cosh({\cal{H}}_1)\Big)e^{ip{\cdot}X} 
\end{aligned}
\label{eq:NS_vertC_-1}
\end{equation}
The one to one correspondence between the mass level and the structure of the corresponding operators is organized as follows, starting from the  explicit form of the mass shell, one can parametrize the increments in mass related to the operators that will appear. Choosing 
\begin{equation}
\begin{split}
&N (g_{n_2},n_2; g_{n_3},n_3) = \sum_{n_2=1}^{\infty} g_{n_2}n_2 + \sum_{n_3=1}^{\infty}g_{n_3}n_3\qquad g_{n_i}\in \mathbb{N}
\end{split}
\label{eq:mass2_A-N}
\end{equation}
\begin{equation}
\begin{split}
&M (l_{m_1},m_1;l_{m_2},m_2) =\begin{cases}
M^{o,e} = \sum_{m_1=1}^{\infty} (2l_{m_1} + 1)\left(m_1 - \frac{1}{2}\right) + \sum_{m_2=1}^{\infty}2l_{m_2}\left(m_2 -\frac{1}{2}\right)\\\\
M^{e,o} = \sum_{m_1=1}^{\infty} 2l_{m_1}\left(m_1 - \frac{1}{2}\right) + \sum_{m_2=1}^{\infty}(2l_{m_2} + 1)\left(m_2 -\frac{1}{2}\right)\qquad \l_{m_i} \in \mathbb{N}\
\end{cases}
\end{split}
\label{eq:mass2_B-M}
\end{equation}
the parameters $n_{i}$ and $g_{n_{i}}$ are related to the operators $({\cal O}_{i})^{g_{n_{i}}}$ while $m_j$ and $l_{m_j}$ to the operators $({\cal H}_{j})^{l_{m_j}}$. In the Tab.\eqref{Tab.1}-\eqref{Tab.3} we display the classification of the operators that compose the vertex operators of the first three mass levels:
\begin{table}[h!]
\hspace{2.1cm}$\bf{\al \mathbb{M}^2 = 0}$

\centering
\begin{tabular}{|c|c|c|c|c|c|c|c|}
\hline
$N$&$M^{o,e}$&$M^{e,o}$&$(g_{n_2},n_2)$&$(g_{n_3},n_3)$&$(l_{m_1},m_1)$&$(l_{m_2},m_2)$&$V^{_{NS}}(z)_{_{(-1)}}$\\
\hline
$0$&$1/2$&$0$&$(0,-)$&$(0,-)$&$(0,1)$&$(0,-)$&${\cal H}_1|_{m_1=1}$\\
\hline
$0$&$0$&$1/2$&$(0,-)$&$(0,-)$&$(0,-)$&$(0,1)$&$-{\cal H}_2|_{m_2=1}$\\
\hline
\end{tabular}
\caption{vertex operator for the massless level}
\label{Tab.1}
\end{table}
\begin{equation}
e^{-\phi}\left( {\cal H}_{1}|_{m_{1}=1} {-} {\cal H}_2|_{m_{2}=1} \right) e^{i p{\cdot}X}=e^{-\phi} \Upsilon_1 {\cdot}\p \,{\cal Q}_{0}^{(1/2,1)} e^{i(p{-}q/2){\cdot}X}- e^{-\phi}q{\cdot}\p\,\Upsilon_1{\cdot}i\pa X\,{\cal Q}_{-1}^{(-1/2,1)}e^{i(p{-}q/2){\cdot}X}
\label{eq:mass2_zero_ex}
\end{equation}
Remembering that ${\cal Q}_{-N}=0$ with $N>0$ and ${\cal Q}_{0}=1$ one finds the vertex operator of the massless vector with polarisation $\Upsilon_1$ and by construction with the physical degrees of freedom only:
\begin{equation}
e^{-\phi}\,\Upsilon_1 {\cdot}\p  \,e^{i(p{-}q/2){\cdot}X}
\label{eq:vertex_Y1Psi}
\end{equation}
For this case we have collected all the possibilities, ignoring for instance that ${\cal H}_2$ gives non-zero contribution if and only if $m_2 >1$. We stress that in the last column, the term $e^{-\phi}e^{ip{\cdot}X}$, is implied. In what follows we collect only the non vanishing contributions.

 From the first massive level Tab.(\ref{Tab.2}) one has the following combination of vertex operators
\begin{table}[h!]
\hspace{1.7cm}$\bf{\al \mathbb{M}^2 = 1}$

\centering
\begin{tabular}{|c|c|c|c|c|c|c|c|}
\hline
$N$&$M^{o,e}$&$M^{e,o}$&$(g_{n_2},n_2)$&$(g_{n_3},n_3)$&$(l_{m_1},m_1)$&$(l_{m_2},m_2)$&$V^{_{NS}}(z)_{_{(-1)}}$\\
\hline
$0$&$3/2$&$0$&$(0,-)$&$(0,-)$&$(0,2)$&$(0,0)$&${\cal H}_1|_{m_1=2}$\\
$0$&$3/2$&$0$&$(0,-)$&$(0,-)$&$(1,1)$&$(0,0)$&${\cal H}^3_1|_{m_1=1}$\\
\hline
$0$&$0$&$3/2$&$(0,-)$&$(0,-)$&$(0,0)$&$(0,2)$&$-{\cal H}_2|_{m_2=2}$\\
\hline
\hline
$1$&$1/2$&$0$&$(1,1)$&$(0,-)$&$(0,1)$&$(0,-)$&${\cal O}_2|_{n_2=1}{\cal H}_1|_{m_1=1}$\\
$1$&$1/2$&$0$&$(0,-)$&$(1,1)$&$(0,1)$&$(0,-)$&${\cal O}_3|_{n_3=1}{\cal H}_1|_{m_1=1}$\\
\hline
\end{tabular}
\caption{vertex operators of the first massive level}
\label{Tab.2}
\end{table}
\begin{equation}
e^{-\phi}\left({\cal H}_1|_{m_{1}=2}{-} {\cal H}_2|_{m_{2}=2}\right)e^{ip{\cdot}X}=e^{-\phi}\left(\Upsilon_2{\cdot}\pa \p - {3\over 2}\Upsilon_2{\cdot}\p\, q{\cdot}i\pa X- q{\cdot}\p\, \Upsilon_2{\cdot}i\pa X \right)e^{i(p{-}3q/2){\cdot}X}
\label{eq:massive1}
\end{equation}
\begin{equation}
e^{-\phi}\left( {\cal O}_2|_{n_{2}=1} {+} {\cal O}_3|_{n_{3}=1}\right)\,{\cal H}_{1}|_{m_{1}=1}\, e^{ip{\cdot}X}= e^{-\phi}\left(\zeta_1{\cdot}i\pa X {-} q{\cdot}\p\,\zeta_{1}{\cdot}\p \right)\,\Upsilon_1{\cdot}\p \, e^{i(p{-}3q/2){\cdot}X}
\label{eq:massive2}
\end{equation}
\begin{equation}
e^{-\phi}({\cal H}_1|_{m_{1}=1})^3 e^{ip{\cdot}X}= e^{-\phi} \Upsilon_1{\cdot}\p \, \Upsilon_1{\cdot}\p\,\Upsilon_1{\cdot}\p\,    e^{i(p{-}3q/2){\cdot}X}
\label{eq:massive3}
\end{equation}
where the total number of the degrees of freedom is $128$, in particular one has 8 d.o.f. from $\Upsilon_2^{\mu}$, 64 d.o.f. from $\zeta_1^{\mu} \Upsilon_1^{\nu}$ and 56 d.o.f. from $\Upsilon_1^{\mu} \Upsilon_1^{\nu} \Upsilon_1^{\rho}$.

\begin{table}[h!]
\hspace{1.2cm}$\bf{\al \mathbb{M}^2 = 2}$

\centering
\begin{tabular}{|c|c|c|c|c|c|c|c|}
\hline
$N$&$M^{o,e}$&$M^{e,o}$&$(g_{n_2},n_2)$&$(g_{n_3},n_3)$&$(l_{m_1},m_1)$&$(l_{m_2},m_2)$&$V^{_{NS}}(z)_{_{(-1)}}$\\
\hline
$0$&$5/2$&$0$&$(0,-)$&$(0,-)$&$(0,3)$&$(0,-)$&${\cal H}_1|_{m_1=3}$\\
$0$&$5/2$&$0$&$(0,-)$&$(0,-)$&$(2,1)$&$(0,-)$&${\cal H}^5_1|_{m_1=1}$\\
\hline
$0$&$0$&$5/2$&$(0,-)$&$(0,-)$&$(0,-)$&$(0,3)$&$-{\cal H}_2|_{m_2=3}$\\
\hline
\hline
$1$&$3/2$&$0$&$(1,1)$&$(0,-)$&$(1,1)$&$(0,-)$&${\cal O}_2|_{n_2=1}{\cal H}^3_1|_{m_1=1}$\\
$1$&$3/2$&$0$&$(1,1)$&$(0,-)$&$(0,2)$&$(0,-)$&${\cal O}_2|_{n_2=1}{\cal H}_1|_{m_1=2}$\\
$1$&$3/2$&$0$&$(0,-)$&$(1,1)$&$(1,1)$&$(0,-)$&${\cal O}_3|_{n_3=1}{\cal H}^3_1|_{m_1=1}$\\
$1$&$3/2$&$0$&$(0,-)$&$(1,1)$&$(0,2)$&$(0,-)$&${\cal O}_3|_{n_3=1}{\cal H}_1|_{m_1=2}$\\
\hline
$1$&$0$&$3/2$&$(1,1)$&$(0,-)$&$(0,-)$&$(0,2)$&-${\cal O}_2|_{n_2=1}{\cal H}_2|_{m_2=2}$\\
$1$&$0$&$3/2$&$(0,-)$&$(1,1)$&$(0,-)$&$(0,2)$&-${\cal O}_3|_{n_3=1}{\cal H}_2|_{m_2=2}$\\
\hline
\hline
$2$&$1/2$&$0$&$(2,1)$&$(0,-)$&$(0,1)$&$(0,-)$&${\cal O}^2_2|_{n_2=1}{\cal H}_1|_{m_1=1}$\\
$2$&$1/2$&$0$&$(1,2)$&$(0,-)$&$(0,1)$&$(0,-)$&${\cal O}_2|_{n_2=2}{\cal H}_1|_{m_1=1}$\\
$2$&$1/2$&$0$&$(0,-)$&$(2,1)$&$(0,1)$&$(0,-)$&${\cal O}^2_3|_{n_3=1}{\cal H}_1|_{m_1=1}$\\
$2$&$1/2$&$0$&$(0,-)$&$(1,2)$&$(0,1)$&$(0,-)$&${\cal O}_3|_{n_3=2}{\cal H}_1|_{m_1=1}$\\
$2$&$1/2$&$0$&$(1,1)$&$(1,1)$&$(0,1)$&$(0,-)$&${\cal O}_2{\cal O}_3|_{n_{\{2,3\}}=1}{\cal H}_1|_{m_1=1}$\\
\hline
\end{tabular}
\caption{vertex operators for the second mass level}
\label{Tab.3}
\end{table}
 Repeating the same procedure of the previous case for the second mass level one can count the d.o.f finding in total 1152 d.o.f. as expected.

\subsection{Extension to the non canonical picture}
To complete the analysis in the NS sector, we proceed and construct explicitly a coherent vertex operator in the first non canonical superghost picture, that is the 0-picture and give the basic ingredients on how to generalize it to the desired picture starting from the canonical one. The 0-picture coherent vertex operator turns out to be necessary to compute string scattering amplitudes with more than two coherent vertex operators  and also needed for the computation of 1-loop couplings (the first non trivial after tree level) involving NS coherent states. 
In order to raise the picture from the canonical to the first non canonical one, we use the picture rising operator $\Pi^{+}$ that acts on the generic operator ${\cal O}_{(\tilde{q})}$, in the picture ($\tilde{q}$), as follows:
\begin{equation}\label{picture action}
\lim_{z\rightarrow w} \Pi^{+}(z) \,{\cal O}_{(\tilde{q})}(w) ={\cal O}_{(\tilde{q}+1)}
\end{equation}
 where, following from \cite{Friedan:1985ge,Friedan:1986rx}, the picture raising operator\footnote{The full form includes terms involving both fermionic and bosonic ghost systems : $-{1 \over 2}\pa\eta e^{2\phi}b - {1\over 2}\pa(\eta e^{2\phi}b)+c\pa\xi$, which we didn't consider because irrelevant at the level of scattering amplitudes \cite{Friedan:1985ge}.} is defined by
 \begin{equation}\label{picture rising}
 \Pi^{+}(z)=-e^{\phi}\,\p{\cdot}i\pa X (z) + ... 
 \end{equation}
 In the case of  coherent vertex operator one has to compute the Non-singular OPE $\Pi^{+}(z)\, {\cal V}^{NS}_{{\cal C}(-1)}(w)$, that explicitly looks
 \begin{equation}\label{OPE rising}
 e^{\phi}\,\p{\cdot}\pa X (z) \,\, e^{-\phi}e^{ {\cal O}_{2}{+}{\cal O}_{3}} \sinh\left( {\cal H}_1{-}{\cal H}_2 \right)   e^{ip{\cdot}X}(w)
 \end{equation}
 One can compute separately the contributions coming from ${\cal O}_2$, ${\cal O}_3$, ${\cal H}_1$ and ${\cal H}_2$ and then use the Wick's rules to implement the results in the full OPE. The first two terms are:
 \begin{equation}\label{rising O2 O3}
 \Pi^{+}(z) \,e^{-\phi}{\cal O}_2(w)|_{z\rightarrow w}={\cal O}_2^{+}(w)\,;\quad \Pi^{+}(z) \,e^{-\phi}{\cal O}_3(w)|_{z\rightarrow w}={\cal O}_3^{+}(w)
 \end{equation}
 where explicit forms of the operators are summarised in Appendix B. To compute these terms a new OPE is needed: the one between $\pa X$ and the cycle index polynomial. It takes
 \begin{equation}\label{DX Z ope}
 i\pa X^{\mu}(z)\, {\cal Z}_N[{\cal U}^{(n)}_{s}](w)= -nq^\mu\sum_{v=1}^{N{-}v}{\cal Z}_{N{-}v}[{\cal U}^{(n)}_{s}](z{-}w)^{-v{-}1} + \sum_{h=1}^{\infty}{i\pa^{h}X^{\mu}\over (h{-}1)!} {\cal Z}_{N}[{\cal U}_{s}^{(n)}] (z{-}w)^{h{-}1}
 \end{equation}
 The last two terms come from:
\begin{equation}\label{rising H2 H3}
 \Pi^{+}(z) \,e^{-\phi}{\cal H}_2(w)|_{z\rightarrow w}={\cal H}_2^{+}(w)\,;\quad \Pi^{+}(z) \,e^{-\phi}{\cal H}_1(w)|_{z\rightarrow w}={\cal H}_1^{+}(w)
 \end{equation} 
 and now the relevant OPE is
 \begin{equation}\label{DX Q ope}
 i\pa X^{\mu}(z)\, {\cal Q}^{(f,m)}_M(w)=-q^{\mu}\sum_{v=1}\left( (m{-}{1\over2}){\cal Q}^{(f,m)}_{M{-}v} {-} v f{\cal Q}_{M{-}v}^{(f-1,m)} \right)(z{-}w)^{-v{-}1} +   \sum_{h=1}^{\infty}{i\pa^{h}X^{\mu}\over (h{-}1)!} {\cal Q}_{M}^{(f,m)} (z{-}w)^{h{-}1}
 \end{equation}
 Implementing this rules and Wick's theorem in (\ref{OPE rising}) one can write down the NS coherent vertex operator in the 0-picture :
 \begin{equation}\label{o-pict coherent}
 {\cal V}^{NS}_{{\cal C}(0)}(z)= e^{{\cal O}_2 {+} {\cal O}_3} \Big[ \left( {\cal O}_2^{+} {+} {\cal O}_3^{+} {+} p{\cdot}\p\right) \sinh\left({\cal H}_1{-}{\cal H}_2\right)  {+}  \left( {\cal H}_1^{+} {-} {\cal H}_2^{+}\right) \cosh\left({\cal H}_1{-}{\cal H}_2\right) \Big] e^{ip{\cdot}X}
 \end{equation}
One can repeat the same analysis as for vertex operator in canonical picture and look at the first three mass levels.
 Now the mass-shell condition involves several combinations due to the presence of ${\cal O}_{2}^{+}, {\cal O}_{3}^{+}, {\cal H}_{1}^{+}$ and ${\cal H}_{2}^{+}$
\begin{equation*}\small{
\begin{aligned}
&N (g_{n_i},n_i;\delta_{n_i},\tilde{n}_i) = \sum_{n_2=1}^{\infty} g_{n_2}n_2 + \sum_{n_3=1}^{\infty}g_{n_3}n_3 + \sum_{\tilde{n}_2=1}^{\infty}\delta_{n_2}\tilde{n}_2 +\sum_{\tilde{n}_3=1}^{\infty}\delta_{n_3}\tilde{n}_3 \qquad g_{n_i}\in \mathbb{N}\,;\,\, \delta_{n_i} \in \{0,1\}
\end{aligned}}
\end{equation*}
\begin{equation}\small{
\begin{aligned}
&M (l_{m_j},m_j;\delta_{m_j},\tilde{m}_j)=\begin{cases}
M^{e} =\sum_{m_1=1}^{\infty} \big(2l_{m_1} + (1 - \delta_{m_1} - \delta_{m_2})\big)\left(m_1 - \frac{1}{2}\right) + \sum_{m_2=1}^{\infty}2l_{m_2}\left(m_2 -\frac{1}{2}\right)\vspace{2mm}\\
\hspace{1.2cm}+ \sum_{\tilde{m}_1=1}^{\infty}\delta_{m_1}\left(\tilde{m}_1 - \frac{1}{2}\right) + \sum_{\tilde{m}_2=1}^{\infty}\delta_{m_2}\left(\tilde{m}_2 - \frac{1}{2}\right)\\\\
M^{o} =\sum_{m_1=1}^{\infty} (2l_{m_1} +\delta_{m_1} + \delta_{m_2} )\left(m_1 - \frac{1}{2}\right) + \sum_{m_2=1}^{\infty}\big(2l_{m_2} + 1\big)\left(m_2 -\frac{1}{2}\right)\vspace{2mm}\\
\hspace{1.2cm}+ \sum_{\tilde{m}_1=1}^{\infty}\delta_{m_1}\left(\tilde{m}_1 - \frac{1}{2}\right) + \sum_{\tilde{m}_2=1}^{\infty}\delta_{m_2}\left(\tilde{m}_2 - \frac{1}{2}\right)\\
\end{cases}
\\
\end{aligned}}
\label{eq:mass2_N,M}
\end{equation}
In the definition of $M$, $M^e$ involves an even number of ${\cal H}_2$ operators while $M^o$ an odd number of $\cal{H}_2$ operators. Moreover the $\delta_{m_i}$ inside $M^{e}$($M^{o}$) is needed, since when terms without ${\cal H}_i^{+}$ operator are considered only odd(even) powers in ${\cal H}_1$ enter, otherwise terms with ${\cal H}_i^{+}$ operator appear only with even(odd)-power in ${\cal H}_1$. All is collected in Tab.\eqref{Tab.4}-\eqref{Tab.6}, and from Tab.\eqref{Tab.5} we will ignore terms involving ${\cal H}_2$ and ${\cal H}^+_2$ unless the non-zero condition $\{m_2,\tilde{m}_2\} >1$ is satisfied.

\begin{table}[h!]
$\bf{\al \mathbb{M}^2 = 0}$

\begin{tabular}{|c|c|c|c|c|c|}
\hline
$N$&$M^{e}$&$M^{o}$&$\begin{aligned}\\{\begin{bmatrix}(g_{n_2},n_2)&\hspace{-3mm}(g_{n_3},n_3)\\(\delta_{n_2},\tilde{n}_2)&\hspace{-3mm}(\delta_{n_3},\tilde{n}_3)\end{bmatrix}}\\\\\end{aligned}$&$\hspace{-1.1cm}{\begin{bmatrix}(l_{m_1},m_1)&\hspace{-3mm}(l_{m_2},m_2)\\(\delta_{m_1},\tilde{m}_1)&\hspace{-3mm}(\delta_{m_2},\tilde{m}_2)\end{bmatrix}}$&$V^{_{NS}}(z)_{_{(0)}}$\\
\hline
$0$&$1/2$&$0$&$\small{\begin{bmatrix}(0,-)&\hspace{-3mm}(0,-)\\(0,-)&\hspace{-3mm}(0,-)\end{bmatrix}}$&$\hspace{-2mm}\begin{aligned}&\small{\begin{bmatrix}(0,1)&\hspace{-3mm}(0,-)\\(0,-)&\hspace{-3mm}(0,-)\end{bmatrix}};\small{\begin{bmatrix}(0,-)&\hspace{-3mm}(0,-)\\(1,1)&\hspace{-3mm}(0,-)\end{bmatrix}}\\&\small{\begin{bmatrix}(0,-)&\hspace{-3mm}(0,-)\\(0,-)&\hspace{-3mm}(1,1)\end{bmatrix}}\end{aligned}$&$\begin{aligned}&p\psi{\cal H}_1|_{m_1=1} + {\cal H}^+_1|_{\tilde{m}_1=1}\\&-{\cal H}^+_2|_{\tilde{m}_2=1}\end{aligned}$\\
\hline
$0$&$0$&$1/2$&$\small{\begin{bmatrix}(0,-)&\hspace{-3mm}(0,-)\\(0,-)&\hspace{-3mm}(0,-)\end{bmatrix}}$&$\small{\begin{bmatrix}(0,-)&\hspace{-3mm}(0,1)\\(0,-)&\hspace{-3mm}(0,-)\end{bmatrix}} $&$- p\psi{\cal H}_2|_{m_2=1}$\\
\hline
\end{tabular}
\caption{zero mass level represented by the coherent operator in the 0 superghost picture}
\label{Tab.4}
\end{table}

\vspace*{-4cm}
\begin{table}[h!]
$\bf{\al \mathbb{M}^2 = 1}$

\begin{tabular}{|c|c|c|c|c|c|}
\hline
$N$&$M^{e}$&$M^{o}$&$\hspace{-2mm}\begin{aligned}\\{\begin{bmatrix}(g_{n_2},n_2)&\hspace{-3mm}(g_{n_3},n_3)\\(\delta_{n_2},\tilde{n}_2)&\hspace{-3mm}(\delta_{n_3},\tilde{n}_3)\end{bmatrix}}\\\\\end{aligned}$&$\hspace{-1.4cm}{\begin{bmatrix}(l_{m_1},m_1)&\hspace{-3mm}(l_{m_2},m_2)\\(\delta_{m_1},\tilde{m}_1)&\hspace{-3mm}(\delta_{m_2},\tilde{m}_2)\end{bmatrix}}$&$V^{_{NS}}(z)_{_{(0)}}$\\
\hline
$0$&$3/2$&$0$&$\small{\begin{bmatrix}(0,-)&\hspace{-3mm}(0,-)\\(0,-)&\hspace{-3mm}(0,-)\end{bmatrix}}$&$\hspace{-2mm}\begin{aligned}&\small{\begin{bmatrix}(0,2)&\hspace{-3mm}(0,-)\\(0,-)&\hspace{-3mm}(0,-)\end{bmatrix}};\small{\begin{bmatrix}(1,1)&\hspace{-3mm}(0,-)\\(0,-)&\hspace{-3mm}(0,-)\end{bmatrix}}\\&\small{\begin{bmatrix}(0,-)&\hspace{-3mm}(0,-)\\(1,2)&\hspace{-3mm}(0,-)\end{bmatrix}};\small{\begin{bmatrix}(0,-)&\hspace{-3mm}(0,-)\\(0,-)&\hspace{-3mm}(1,2)\end{bmatrix}}\\&\small{\begin{bmatrix}(1,1)&\hspace{-3mm}(0,-)\\(1,1)&\hspace{-3mm}(0,-)\end{bmatrix}}\end{aligned}$&$\begin{aligned}&p\psi{\cal H}_1|_{m_1=2} + p\psi{\cal H}^3_1|_{m_1=1}\\&+{\cal H}^+_1|_{\tilde{m}_1=2}-{\cal H}^+_2|_{\tilde{m}_2=2}\\&+{\cal H}^+_1{\cal H}^2_1|_{\{m_1,\tilde{m}_1\}=1}\end{aligned}$\\
\hline
$0$&$0$&$3/2$&$\small{\begin{bmatrix}(0,-)&\hspace{-3mm}(0,-)\\(0,-)&\hspace{-3mm}(0,-)\end{bmatrix}}$&$\small{\begin{bmatrix}(0,-)&\hspace{-3mm}(0,2)\\(0,-)&\hspace{-3mm}(0,-)\end{bmatrix}}$&$-p\psi{\cal H}_2|_{m_2=2}$\\
\hline
\hline
$1$&$1/2$&$0$&$\begin{aligned}&\small{\begin{bmatrix}(1,1)&\hspace{-3mm}(0,-)\\(0,-)&\hspace{-3mm}(0,-)\end{bmatrix}}\\&\small{\begin{bmatrix}(0,-)&\hspace{-3mm}(1,1)\\(0,-)&\hspace{-3mm}(0,-)\end{bmatrix}}\end{aligned}$&$\hspace{-3mm}\begin{aligned}&\small{\begin{bmatrix}(0,1)&\hspace{-3mm}(0,-)\\(0,-)&\hspace{-3mm}(0,-)\end{bmatrix}};\small{\begin{bmatrix}(0,-)&\hspace{-3mm}(0,-)\\(1,1)&\hspace{-3mm}(0,-)\end{bmatrix}}\end{aligned}$&$\begin{aligned}&{\cal O}_2|_{n_2=1}(p\psi{\cal H}_1|_{{m}_1=1}+{\cal H}^+_1|_{\tilde{m}_1=1})\\&{\cal O}_3|_{n_3=1}(p\psi{\cal H}_1|_{{m}_1=1}+{\cal H}^+_1|_{\tilde{m}_1=1})\end{aligned}$\\
\hline
$1$&$1/2$&$0$&$\begin{aligned}&\small{\begin{bmatrix}(0,-)&\hspace{-3mm}(0,-)\\(1,1)&\hspace{-3mm}(0,-)\end{bmatrix}}\\&\small{\begin{bmatrix}(0,-)&\hspace{-3mm}(0,-)\\(0,-)&\hspace{-3mm}(1,1)\end{bmatrix}}\end{aligned}$&$\small{\begin{bmatrix}(0,1)&\hspace{-3mm}(0,-)\\(0,-)&\hspace{-3mm}(0,-)\end{bmatrix}}$&$({\cal O}^+_2|_{\tilde{n}_2=1} + {\cal O}^+_3|_{\tilde{n}_3=1}){\cal H}_1|_{m_1=1}$\\
\hline
\end{tabular}
\caption{first massive level represented by the coherent operator in the 0 superghost picture.}
\label{Tab.5}
\end{table}
\vspace{-7cm}
\begin{table}[h!]
$\bf{\al \mathbb{M}^2 = 2}\,\,(Part I)$

\footnotesize{
\begin{tabular}{|c|c|c|c|c|c|c|}
\hline
$\hspace{-2mm}N$&$\hspace{-2mm}M^{e}$&$\hspace{-2mm}M^{o}$&$\hspace{-1mm}{\begin{bmatrix}(g_{n_2},n_2)&\hspace{-3mm}(g_{n_3},n_3)\\(\delta_{n_2},\tilde{n}_2)&\hspace{-3mm}(\delta_{n_3},\tilde{n}_3)\end{bmatrix}}$&${\begin{bmatrix}(l_{m_1},m_1)&\hspace{-3mm}(l_{m_2},m_2)\\(\delta_{m_1},\tilde{m}_1)&\hspace{-3mm}(\delta_{m_2},\tilde{m}_2)\end{bmatrix}}$&$V^{_{NS}}(z)_{_{(0)}}$\\
\hline
$0$&$5/2$&$0$&$\hspace{-5mm}\begin{aligned}&\small{\begin{bmatrix}(0,-)&\hspace{-3mm}(0,-)\\(0,-)&\hspace{-3mm}(0,-)\end{bmatrix}}\end{aligned}$&$\hspace{-3mm}\begin{aligned}&\small{\begin{bmatrix}(0,3)&\hspace{-3mm}(0,-)\\(0,-)&\hspace{-3mm}(0,-)\end{bmatrix}};\small{\begin{bmatrix}(2,1)&\hspace{-3mm}(0,-)\\(0,-)&\hspace{-3mm}(0,-)\end{bmatrix}}\\&\small{\begin{bmatrix}(0,-)&\hspace{-3mm}(0,-)\\(1,3)&\hspace{-3mm}(0,-)\end{bmatrix}};\small{\begin{bmatrix}(0,-)&\hspace{-3mm}(0,-)\\(0,-)&\hspace{-3mm}(1,3)\end{bmatrix}}\\&\small{\begin{bmatrix}(1,1)&\hspace{-3mm}(0,-)\\(1,2)&\hspace{-3mm}(0,-)\end{bmatrix}};\small{\begin{bmatrix}(1,1)&\hspace{-3mm}(0,-)\\(0,-)&\hspace{-3mm}(1,2)\end{bmatrix}}\\&\small{\begin{bmatrix}(2,1)&\hspace{-3mm}(0,-)\\(1,1)&\hspace{-3mm}(0,-)\end{bmatrix}}\end{aligned}$&$\begin{aligned}&p\psi{\cal H}_1|_{m_1=3} +p\psi{\cal H}^5_1|_{m_1=1}\\&+{\cal H}^+_1|_{m_1=3} - {\cal H}^+_2|_{m_2=3}\\& + {\cal H}_1|_{m_1=1}({\cal H}^+_1|_{\tilde{m}_1=2} - {\cal H}^+_2|_{\tilde{m}_2=2}) \\&+{\cal H}^+_1{\cal H}^4_1|_{\{\tilde{m}_1,m_1\}=1}\end{aligned}$\\
\hline
$0$&$0$&$5/2$&$\hspace{-5mm}\begin{aligned}&\small{\begin{bmatrix}(0,-)&\hspace{-3mm}(0,-)\\(0,-)&\hspace{-3mm}(0,-)\end{bmatrix}}\end{aligned}$&$\hspace{-3mm}\begin{aligned}&\small{\begin{bmatrix}(0,-)&\hspace{-3mm}(0,3)\\(0,-)&\hspace{-3mm}(0,-)\end{bmatrix}};\small{\begin{bmatrix}(1,1)&\hspace{-3mm}(0,2)\\(0,-)&\hspace{-3mm}(0,-)\end{bmatrix}}\\&\small{\begin{bmatrix}(0,1)&\hspace{-3mm}(0,2)\\(1,1)&\hspace{-3mm}(0,-)\end{bmatrix}}\end{aligned}$&$ \begin{aligned}&- p\psi{\cal H}_2|_{m_2=3} -p\psi{\cal H}^2_1|_{m_1=1}{\cal H}_2|_{m_2=2}\\&-{\cal H}^+_1{\cal H}_1|_{\{\tilde{m}_1,m_1\}=1}{\cal H}_2|_{m_2=2}\end{aligned} $\\
\hline
\hline
$1$&$3/2$&$0$&$\hspace{-8mm}\begin{aligned}&\small{\begin{bmatrix}(1,1)&\hspace{-3mm}(0,-)\\(0,-)&\hspace{-3mm}(0,-)\end{bmatrix}}\\&\small{\begin{bmatrix}(0,-)&\hspace{-3mm}(1,1)\\(0,-)&\hspace{-3mm}(0,-)\end{bmatrix}}\end{aligned}$&$\hspace{-3mm}\begin{aligned}&\small{\begin{bmatrix}(0,2)&\hspace{-3mm}(0,-)\\(0,-)&\hspace{-3mm}(0,-)\end{bmatrix}};\small{\begin{bmatrix}(1,1)&\hspace{-3mm}(0,-)\\(0,-)&\hspace{-3mm}(0,-)\end{bmatrix}}\\&\small{\begin{bmatrix}(0,-)&\hspace{-3mm}(0,-)\\(1,2)&\hspace{-3mm}(0,-)\end{bmatrix}};\small{\begin{bmatrix}(0,-)&\hspace{-3mm}(0,-)\\(0,-)&\hspace{-3mm}(1,2)\end{bmatrix}}\\&\small{\begin{bmatrix}(1,1)&\hspace{-3mm}(0,-)\\(1,1)&\hspace{-3mm}(0,-)\end{bmatrix}}\end{aligned}$&$\hspace{-1.9mm}({\cal O}_2 {+} {\cal O}_3)|_{\{n_2,n_3\}=1}\left(\begin{aligned}&p\psi\left({\cal H}_1|_{m_1=2} {+} {\cal H}^3_1|_{m_1=1}\right)\\&{\cal H}^+_1|_{\tilde{m}_1=2}{-}{\cal H}^+_2|_{\tilde{m}_2=2}\\&{\cal H}^+_1{\cal H}^2_1|_{\{m_1,\tilde{m}_1\}=1}\end{aligned}\right)$\\
\hline
\end{tabular}}
\caption{second massive level represented by the coherent operator in the 0 superghost picture.}
\end{table}

\newpage
\begin{table}[h!]
$\bf{\al \mathbb{M}^2 = 2}\,\,(Part II)$

\footnotesize{
\begin{tabular}{|c|c|c|c|c|c|c|}
\hline
$\hspace{-2mm}N$&$\hspace{-2mm}M^{e}$&$\hspace{-2mm}M^{o}$&$\hspace{-1mm}{\begin{bmatrix}(g_{n_2},n_2)&\hspace{-3mm}(g_{n_3},n_3)\\(\delta_{n_2},\tilde{n}_2)&\hspace{-3mm}(\delta_{n_3},\tilde{n}_3)\end{bmatrix}}$&${\begin{bmatrix}(l_{m_1},m_1)&\hspace{-3mm}(l_{m_2},m_2)\\(\delta_{m_1},\tilde{m}_1)&\hspace{-3mm}(\delta_{m_2},\tilde{m}_2)\end{bmatrix}}$&$V^{_{NS}}(z)_{_{(0)}}$\\
\hline
$1$&$3/2$&$0$&$\hspace{-8mm}\begin{aligned}&\small{\begin{bmatrix}(0,-)&\hspace{-3mm}(0,-)\\(1,1)&\hspace{-3mm}(0,-)\end{bmatrix}}\\&\small{\begin{bmatrix}(0,-)&\hspace{-3mm}(0,-)\\(0,-)&\hspace{-3mm}(1,1)\end{bmatrix}}\end{aligned}$&$\hspace{-2mm}\small{\begin{bmatrix}(0,2)&\hspace{-3mm}(0,-)\\(0,-)&\hspace{-3mm}(0,-)\end{bmatrix}};\small{\begin{bmatrix}(1,1)&\hspace{-3mm}(0,-)\\(0,-)&\hspace{-3mm}(0,-)\end{bmatrix}}$&$({\cal O}^+_2+{\cal O}^+_3)|_{\{\tilde{n}_2,\tilde{n}_3\}=1}\left({\cal H}_1|_{m_1=2} + {\cal H}^3_1|_{m_1=1}\right)$\\
\hline
$1$&$0$&$3/2$&$\hspace{-8mm}\begin{aligned}&\small{\begin{bmatrix}(1,1)&\hspace{-3mm}(0,-)\\(0,0)&\hspace{-3mm}(0,-)\end{bmatrix}}\\&\small{\begin{bmatrix}(0,-)&\hspace{-3mm}(1,1)\\(0,0)&\hspace{-3mm}(0,-)\end{bmatrix}}\end{aligned}$&$\small{\begin{bmatrix}(0,-)&\hspace{-3mm}(0,2)\\(0,-)&\hspace{-3mm}(0,-)\end{bmatrix}}$&$-({\cal O}_2+{\cal O}_3)|_{\{n_2,n_3\}=1}p\psi{\cal H}_2|_{m_2=2}$\\
\hline
$1$&$0$&$3/2$&$\hspace{-8mm}\begin{aligned}&\small{\begin{bmatrix}(0,-)&\hspace{-3mm}(0,-)\\(1,1)&\hspace{-3mm}(0,-)\end{bmatrix}}\\&\small{\begin{bmatrix}(0,-)&\hspace{-3mm}(0,-)\\(0,0)&\hspace{-3mm}(1,1)\end{bmatrix}}\end{aligned}$&$\small{\begin{bmatrix}(0,-)&\hspace{-3mm}(0,2)\\(0,-)&\hspace{-3mm}(0,-)\end{bmatrix}}$&$-({\cal O}^+_2+{\cal O}^+_3)|_{\{\tilde{n}_2,\tilde{n}_3\}=1}{\cal H}_2|_{m_2=2}$\\
\hline
\hline
$2$&$1/2$&$0$&$\hspace{-8mm}\begin{aligned}&\small{\begin{bmatrix}(2,1)&\hspace{-3mm}(0,-)\\(0,0)&\hspace{-3mm}(0,-)\end{bmatrix}}\\&\small{\begin{bmatrix}(0,-)&\hspace{-3mm}(2,1)\\(0,0)&\hspace{-3mm}(0,-)\end{bmatrix}}\\&\small{\begin{bmatrix}(1,2)&\hspace{-3mm}(0,-)\\(0,0)&\hspace{-3mm}(0,-)\end{bmatrix}}\\&\small{\begin{bmatrix}(0,-)&\hspace{-3mm}(1,2)\\(0,0)&\hspace{-3mm}(0,-)\end{bmatrix}}\end{aligned}$&$\hspace{-2mm}\small{\begin{bmatrix}(0,1)&\hspace{-3mm}(0,-)\\(0,-)&\hspace{-3mm}(0,-)\end{bmatrix}};\small{\begin{bmatrix}(0,-)&\hspace{-3mm}(0,-)\\(1,1)&\hspace{-3mm}(0,-)\end{bmatrix}}$&$\hspace{-2mm}\left(\begin{aligned}&({\cal O}^2_2{+} {\cal O}^2_3)|_{\{n_2,n_3\}=1}\\&({\cal O}_2{+} {\cal O}_3)|_{\{n_2,n_3\}=2}\end{aligned}\right)({\cal H}^+_1|_{\tilde{m}_1=1} {+} p\psi{\cal H}_1|_{{m}_1=1})$\\
\hline
$2$&$1/2$&$0$&$\hspace{-8mm}\begin{aligned}&\small{\begin{bmatrix}(0,-)&\hspace{-3mm}(0,-)\\(1,2)&\hspace{-3mm}(0,-)\end{bmatrix}}\\&\small{\begin{bmatrix}(0,-)&\hspace{-3mm}(0,-)\\(0,0)&\hspace{-3mm}(1,2)\end{bmatrix}}\end{aligned}$&$\small{\begin{bmatrix}(0,1)&\hspace{-3mm}(0,-)\\(0,-)&\hspace{-3mm}(0,-)\end{bmatrix}}$&$({\cal O}^+_2+{\cal O}^+_3)|_{\{\tilde{n}_2,\tilde{n}_3\}=2}{\cal H}_1|_{{m}_1=1}$\\
\hline
$2$&$1/2$&$0$&$\hspace{-5.5mm}\begin{aligned}&\small{\begin{bmatrix}(1,1)&\hspace{-3mm}(1,1)\\(0,0)&\hspace{-3mm}(0,-)\end{bmatrix}}\end{aligned}$&$\hspace{-2mm}\small{\begin{bmatrix}(0,1)&\hspace{-3mm}(0,-)\\(0,-)&\hspace{-3mm}(0,-)\end{bmatrix}};\small{\begin{bmatrix}(0,-)&\hspace{-3mm}(0,-)\\(1,1)&\hspace{-3mm}(0,-)\end{bmatrix}}$&${\cal O}_2{\cal O}_3|_{\{n_2,n_3\}=1}({\cal H}^+_1|_{\tilde{m}_1=1} {+} p\psi{\cal H}_1|_{{m}_1=1})$\\
\hline
$2$&$1/2$&$0$&$\hspace{-5.5mm}\begin{aligned}&\small{\begin{bmatrix}(1,1)&\hspace{-3mm}(1,1)\\(1,1)&\hspace{-3mm}(1,1)\end{bmatrix}}\end{aligned}$&$\small{\begin{bmatrix}(0,1)&\hspace{-3mm}(0,-)\\(0,-)&\hspace{-3mm}(0,-)\end{bmatrix}}$&$\hspace{-2mm}({\cal O}_2{+}{\cal O}_3)({\cal O}^+_2{+}{\cal O}^+_3)|_{\{n_2,n_3,\tilde{n}_2,\tilde{n}_3\}=1}{\cal H}_1|_{m_1=1}$\\
\hline
\end{tabular}}
\caption{second massive level represented by the coherent operator in the 0 superghost picture.}
\label{Tab.6}
\end{table}

\newpage
 \section{Three-points Amplitudes}
  In this section we will discuss the role of the coherent vertex operator in scattering amplitudes processes. To this end we compute the simplest three-points coupling involving a NS coherent state in two different pictures.
 \subsection{Coherent state-Vector-Vector coupling (canonical picture)}
Considering the coherent vertex operator of (\ref{simple NS coher}), we want to compute the three-points disk coupling between a NS coherent state and two vectors, that is 
\begin{equation}\label{3pt coupling}
C_{3}\left( \{\zeta_n\},\{\Upsilon_m\},p,q\,;A_2,k_2\,; A_3,k_3 \right)=\left\langle c{\cal V}^{NS}_{{\cal C}\,(-1)}(z_1)\, c{\cal V}_{A\,(-1)}(z_2)\, c{\cal V}_{A\,(0)}(z_3)\right\rangle
\end{equation}
in which we insert the c-ghosts in order to fix the $PSL(2,\mathbb{R})$ symmetry. Barring the Chan-Paton factors the vector vertex operators read
\begin{equation}\label{vector vertex operators}
{\cal V}_{A\,(-1)}(z)=e^{-\phi} A{\cdot}\p\, e^{ik{\cdot}X}(z) \,; \quad {\cal V}_{A\,(0)}(z)= \left( A{\cdot}i\pa X+ k{\cdot}\p A{\cdot}\p \right) \,e^{ik{\cdot}X}(z) \,; \quad k^2=A{\cdot}k=0
\end{equation}
 First of all one can factor out the ghosts and super ghosts contributions
\begin{equation}
 \left\langle c(z_1) c(z_2) c(z_3) \right\rangle =z_{12}z_{13}z_{23} \,;\quad  \left\langle e^{-\phi}(z_1) e^{-\phi}(z_2)  \right\rangle=z_{12}^{-1}
 \label{eq:c-gost,s-ghost}
 \end{equation} 
 where $z_{i j}=z_i-z_j$, and evaluate the main contribution
 \begin{equation}\label{amp explit pict -1}
  \left\langle  e^{ {\cal O}_{2}{+}{\cal O}_{3}} \sinh \left( {\cal H}_{1} {-} {\cal H}_{2} \right) e^{ip{\cdot}X}(z_1)A_2{\cdot}\p e^{ik_2{\cdot}X}(z_2)\left( A_3{\cdot}i\pa X{+} k_3{\cdot}\p A_3{\cdot}\p \right) e^{ik_3{\cdot}X}(z_3) \right\rangle
 \end{equation}
 Remembering that ${\cal O}_i$ operators and ${\cal H}_j$ operators are referred respectively to worldsheet bosons and fermions modes, in so far as the light like momentum is concerned one has $e^{-inq{\cdot}X}$ insertions from ${\cal O}_i$ and $e^{-i(m{-}1/2)q{\cdot}X}$ from the others. Using this fact one can write the Koba-Nielsen (KN) factor as:
 \begin{equation}\label{KN}
 \begin{split}
& \left\langle e^{i p_{{\cal C}}{\cdot}X}(z_1) \,e^{i k_2{\cdot}X}(z_2)\,e^{i k_3{\cdot}X}(z_3)\right\rangle=z_{12}^{p_{\cal C}{\cdot}k_2}  z_{13}^{p_{\cal C}{\cdot}k_3} z_{23}^{k_2{\cdot}k_3}\,(2\pi)^D\delta\left(p_{{\cal C}}+k_2+k_3 \right)\\
 &=\int_{{\cal M}_{D}} d^D X_0 \,e^{i(p+k_2+k_3){\cdot}X_0}\,e^{-i(N+M)q{\cdot}X_0} z_{12}^{-1{+}N{+}M}  \,z_{13}^{-1{+}N{+}M} z_{23}^{1{-}N{-}M}
 \end{split}
 \end{equation}
where $p^\mu_{{\cal C}}=p^\mu{-}(N+M)q^\mu$ and $N$, $M$ take into account the light-like momentum insertions according to (\ref{Mddf}) extending the mass shell composition to the case of coherent superposition, the general form of $N$ is:
\begin{equation}
N= \sum_{n_2=1}  g_{n_2} n_2 + \sum_{n_3=1} n_3\,;\quad g_{n_2}\in \mathbb{N}
\label{eq:N_mass2}
\end{equation}

are related, neatly,  to ${\cal O}_2$ and ${\cal O}_3$ and the coefficient $g_{n_2}$ come from the exponential structure of ${\cal O}_2$ , while for $M$ :
\begin{equation}
M= \begin{cases}\sum_{m_2=1} \ell_{m_2} \left(m_2{-}{1\over 2}\right) + \sum_{m_1=1}\left({m_1{-}{1\over 2}}\right)  \,; \quad  {\ell}_{m_2}\,\, \text{even}\\\\ \sum_{m_2=1} \ell_{m_2} \left(m_2{-}{1\over 2}\right) \,; \hspace*{3.7cm} {\ell}_{m_2}\,\, \text{odd}  \end{cases}
\label{eq:M_mass2}
\end{equation}
 where $m_1$ is related with ${\cal H}_1$, $m_2$ with ${\cal H}_2$ and $\ell_{m_{2}}$ a positive integer even or odd depending on the hyperbolic cosine or sine structure. 
Looking at the sub-amplitude structures, identified by the number of fermionic contractions and the relations $q{\cdot}A_{2,3}{=}0$, one has in total six contributions $S_1$,$S_2$,$S_3$,$S_4$,$S_5$,$S_6$ that have in common the integral of momentum conservation and the contribution coming from $e^{{\cal O}_2}$ that is :
\begin{equation}\label{coomon factor}
\int_{{\cal M}_D} d^D X_0 \, e^{i(p{+}k_2{+}k_3){\cdot}X_0} \exp\left(\sum_{n_{2}=1} {(-)^{n_{2}{+}1}\over 2\,n_{2}!}  {\Gamma\left[ {n_{2}\over 2}(Q{+}1) \right] \over \Gamma\left[{n_{2}\over 2}(Q{-}1){+}1) \right]}\,\widetilde{\zeta}_{n_{2}}{\cdot}(k_2{-}k_{3})\right) = \int_{{\cal M}_D} dX_0 \, e^{F(X_0)}
\end{equation}
that for simplicity we abbreviate as above. With this redefinition and the fact that the polarisations of the coherent state, after the contractions,  get ``dressed" according to
\begin{equation}
   \widetilde{\zeta}_{n}=\zeta_{n}\, e^{-inq{\cdot}X_{0}}\,;\quad \widetilde{\Upsilon}_{m}=\Upsilon_{m}\,e^{-i(m{-}1/2)q{\cdot}X_0}
\label{eq:dress_pol}
\end{equation}   
the sub-amplitudes take the following form:
\begin{equation}\label{S1}
\begin{split}
S_1&= \left\langle c(z_1) c(z_2) c(z_3) \right\rangle\, \left\langle e^{-\phi}(z_1) e^{-\phi}(z_2)  \right\rangle\,\left\langle e^{{\cal O}_2}{\cal H}_{1}\,e^{ip{\cdot}X}(z_1) \,A_2{\cdot}\p e^{ik_2{\cdot}X}(z_2)\, A_3{\cdot}i\pa X_3 e^{ik_3{\cdot}X}(z_3)  \right\rangle \\
&=\int_{{\cal M}_D} d^D X_0 \, e^{F(X_0)} \sum_{m_{1}=1}{(-)^{m_{1}{+}1}\over 2^{m_{1}{-}2}}\sum_{n_{2}=1} \Upsilon_{m_{1}}{\cdot}A_2\,\zeta_{n_{2}}{\cdot}A_3\,  {\Gamma[1{+}{n_2 \over 2}(1{+}Q)] \over n_2!\,\Gamma[2{+}{n_2 \over 2}(Q{-}1)]} \sum_{\ell=0} {\cal Q}^{(1/2,m_{1})}_{m_{1}{-}\ell{-}1} \,+\\
&+ \int_{{\cal M}_D} d^D X_0 \, e^{F(X_0)} \sum_{m_{1}=1}{(-)^{m_{1}}\over 2^{m_{1}{-}1}}\widetilde{\Upsilon}_{m_{1}}{\cdot}A_{2}\, A_{3}{\cdot}(k_2{-}p_{{\cal C}})\sum_{\ell=0} {\cal Q}^{(1/2,m_{1})}_{m_{1}{-}\ell{-}1} 
\end{split}
\end{equation}
\begin{equation}
\begin{split}
S_2&= \left\langle c(z_1) c(z_2) c(z_3) \right\rangle\, \left\langle e^{-\phi}(z_1) e^{-\phi}(z_2)  \right\rangle\,\left\langle e^{{\cal O}_2}{\cal O}_3{\cal H}_{1}\,e^{ip{\cdot}X}(z_1) \,A_2{\cdot}\p e^{ik_2{\cdot}X}(z_2)\, k_3{\cdot}\p A_3{\cdot}\p e^{ik_3{\cdot}X}(z_3)   \right\rangle \\
&=\int_{{\cal M}_D} d^D X_0 \, e^{F(X_0)} \sum_{n_{3}=1} {(Q{+}1)\over (n_{3}{-}1)!} {\Gamma\left[ {n_{3}\over2}(1{-}Q) \right] \over \Gamma\left[1{-}{n_{3}\over 2}(1{+}Q)\right]}\\
&\hspace{0.5cm} \sum_{m_{1}=1}{(-)^{m_{4}{+}1}\over 2^{m_{4}{+}1}} \left( { {n_{3}\over 2}(1{-}Q)  \over 1{-}{n_{3}\over 2}(1{+}Q)}\widetilde{\zeta}_{n_{3}}{\cdot}A_{3}\,\widetilde{\Upsilon}_{m_{4}}{\cdot}A_2-\widetilde{\zeta}_{n_{3}}{\cdot}A_2\,\widetilde{\Upsilon}_{m_{4}}{\cdot}A_3 \right) \sum_{h=0}{\cal Q}_{m_{4}{-}h{-}1}^{(1/2,m_{4})}
\end{split}
\label{eq:S2}
\end{equation}
\begin{equation}
\begin{split}
S_3&=-\left\langle c(z_1) c(z_2) c(z_3) \right\rangle\, \left\langle e^{-\phi}(z_1) e^{-\phi}(z_2)  \right\rangle\,\left\langle e^{{\cal O}_2}{\cal H}_{2}\,e^{ip{\cdot}X}(z_1) \,A_2{\cdot}\p e^{ik_2{\cdot}X}(z_2)\, k_3{\cdot}\p A_3{\cdot}\p e^{ik_3{\cdot}X}(z_3)   \right\rangle \\
&=\int_{{\cal M}_D} d^D X_0 \, e^{F(X_0)}\sum_{m_{2}=1}{(Q{+}1)\over 2^{m_{2}{+}2}}A_{2}{\cdot}A_3\,\widetilde{\Upsilon}_{m_{2}}{\cdot}(k_3{-}k_2) \sum_{h_{1},h_{2}=0} (-)^{m_2{+}h_{1}{+}1} \Big( (-)^{h_{2}}{-}1\Big) {\cal Q}_{m_{2}{-}h_{1}{-}h_{2}{-}1}^{(-1/2,m_{2})}
\end{split}
\label{eq:S3}
\end{equation}
\begin{equation}
\begin{split}
S_4&=\left\langle c(z_1) c(z_2) c(z_3) \right\rangle\, \left\langle e^{-\phi}(z_1) e^{-\phi}(z_2)\right\rangle\,\left\langle e^{{\cal O}_2}{\cal H}_{1}\,e^{ip{\cdot}X}(z_1) \,A_2{\cdot}\p e^{ik_2{\cdot}X}(z_2)\, k_3{\cdot}\p A_3{\cdot}\p e^{ik_3{\cdot}X}(z_3)   \right\rangle  \\
&=\int_{{\cal M}_D} d^D X_0 \, e^{F(X_0)}\sum_{m_{1}=1}{(-)^{m_{1}{+}1}\over 2^{m_{1}}}\left( A_2{\cdot}(k_3{-}p_{{\cal C}})\,\widetilde{\Upsilon}_{m_{1}}{\cdot}A_{3} - A_{2}{\cdot}A_3\,\widetilde{\Upsilon}_{m_{1}}{\cdot}(k_3{-}k_2)   \right)  \sum_{h=0} (-)^{h} {\cal Q}_{m_{1}{-}h{-}1}^{(1/2,m_{4})}
\end{split}
\label{eq:S4}
\end{equation}
\begin{equation}
\begin{split}
S_5&=\left\langle c(z_1) c(z_2) c(z_3) \right\rangle\, \left\langle e^{-\phi}(z_1) e^{-\phi}(z_2)\right\rangle\,\left\langle e^{{\cal O}_2}({\cal H}_{1})^3\,e^{ip{\cdot}X}(z_1) \,A_2{\cdot}\p e^{ik_2{\cdot}X}(z_2)\, k_3{\cdot}\p A_3{\cdot}\p e^{ik_3{\cdot}X}(z_3)   \right\rangle  \\
&=\int_{{\cal M}_D}d^DX_0 \, e^{F(X_0)}\sum_{m_{1},m_{2},m_{3}=1}  \sum_{h_{1},h_{2},h_{3}=0}   {(-)^{m_{1}{+}m_2{+}m_3}\over 2^{m_{1}{+}m_2{+}m_3{-}3}} {\cal Q}_{m_{1}{-}h_1{-}1}^{(1/2,m_{1})} {\cal Q}_{m_{2}{-}h_2{-}1}^{(1/2,m_{2})}  {\cal Q}_{m_{3}{-}h_3{-}1}^{(1/2,m_{3})} (-)^{h_1{+}h_2{+}h_3} \\
&\big[(-)^{h_2} \widetilde{\Upsilon}_{m_1}{\cdot}A_3\widetilde{\Upsilon}_{m_2}{\cdot}A_2\widetilde{\Upsilon}_{m_3}{\cdot}k_3 - (-)^{h_3}\widetilde{\Upsilon}_{m_1}{\cdot}A_3\widetilde{\Upsilon}_{m_2}{\cdot}k_3\widetilde{\Upsilon}_{m_3}{\cdot}A_2 +(-)^{h_2}\widetilde{\Upsilon}_{m_3}{\cdot}k_3\widetilde{\Upsilon}_{m_1}{\cdot}A_3\widetilde{\Upsilon}_{m_2}{\cdot}A_2+\\
&-(-)^{h_2}\widetilde{\Upsilon}_{m_1}{\cdot}k_3\widetilde{\Upsilon}_{m_2}{\cdot}A_2\widetilde{\Upsilon}_{m_3}{\cdot}A_3 + (-)^{h_2}\widetilde{\Upsilon}_{m_1}{\cdot}k_3\widetilde{\Upsilon}_{m_2}{\cdot}A_3\widetilde{\Upsilon}_{m_3}{\cdot}A_2 + (-)^{h_1}\widetilde{\Upsilon}_{m_2}{\cdot}A_3\widetilde{\Upsilon}_{m_1}{\cdot}A_2\widetilde{\Upsilon}_{m_3}{\cdot}k_3+\\
& - (-)^{h_3}\widetilde{\Upsilon}_{m_2}{\cdot}A_3\widetilde{\Upsilon}_{m_1}{\cdot}k_3\widetilde{\Upsilon}_{m_3}{\cdot}A_2 -
 (-)^{h_1}\widetilde{\Upsilon}_{m_2}{\cdot}k_3\widetilde{\Upsilon}_{m_1}{\cdot}A_2\widetilde{\Upsilon}_{m_3}{\cdot}A_3 + (-)^{h_3}\widetilde{\Upsilon}_{m_2}{\cdot}k_3\widetilde{\Upsilon}_{m_1}{\cdot}A_3\widetilde{\Upsilon}_{m_3}{\cdot}A_2+\\
 &+(-)^{h_1}\Upsilon_{m_3}{\cdot}A_3\Upsilon_{m_1}{\cdot}A_2\Upsilon_{m_1}{\cdot}k_3 - (-)^{h_1}\Upsilon_{m_3}{\cdot}A_3\Upsilon_{m_1}{\cdot}k_3\Upsilon_{m_2}{\cdot}A_2 -(-)^{h_1}\Upsilon_{m_3}{\cdot}k_3\Upsilon_{m_1}{\cdot}A_2\Upsilon_{m_2}{\cdot}A_3 \big]\\
\end{split}
\label{eq:S5}
\end{equation}
\begin{equation}
\begin{split}
S_6&= \left\langle c(z_1) c(z_2) c(z_3) \right\rangle\, \left\langle e^{-\phi}(z_1) e^{-\phi}(z_2)  \right\rangle\,\left\langle e^{{\cal O}_2}({\cal H}_{1})^2 {\cal H}_2\,e^{ip{\cdot}X}(z_1) \,A_2{\cdot}\p e^{ik_2{\cdot}X}(z_2)\, k_3{\cdot}\p A_3{\cdot}\p e^{ik_3{\cdot}X}(z_3)   \right\rangle \\
&=\int_{{\cal M}_D}d^DX_0 \, e^{F(X_0)}  \sum_{m_{1},m_{2},m_{3}=1}  \sum_{h_{1},h_{2},h_{3},h_{4}=0}   {(-)^{m_{1}{+}m_2{+}m_3}\over 2^{m_{1}{+}m_2{+}m_3{-}2}} {\cal Q}_{m_{1}{-}h_1{-}1}^{(1/2,m_{1})} {\cal Q}_{m_{2}{-}h_2{-}1}^{(1/2,m_{2})}  {\cal Q}_{m_{3}{-}h_3{-}h_4{-}1}^{(-1/2,m_{3})} \\
&(-)^{h_1{+}h_2{+}h_3{+}h_4}\Big(1-(-)^{h_4} \Big)\,(Q{+}1)\,\widetilde{\Upsilon}_{m_{3}}{\cdot}k_3\, \Big( (-)^{h_1}\widetilde{\Upsilon}_{m_{1}}{\cdot}A_2\,\widetilde{\Upsilon}_{m_{2}}{\cdot}A_3 + (-)^{h_2} \widetilde{\Upsilon}_{m_{2}}{\cdot}A_2\, \widetilde{\Upsilon}_{m_{1}}{\cdot}A_3 \Big)
\end{split}
\label{S6}
\end{equation}
where the argument inside the ${\cal Q}$ polynomial takes the following form:
\begin{equation}
{\cal U}_{s}^{(\ell)}={\ell \over 2} \Big( Q - 1 + (-)^{s{+1}}(Q+1) \Big)
\label{eq:arg_Q}
\end{equation}
All the details of the various summations are reported in the Appendix A. It's very easy to select the three vector coupling, looking at $\alpha'\mathbb{(M)}^2 = \left( N{+}M{-}1\right)$, one needs $N=0$ and $M=1$ that means no insertions of $n$ and only one of type $m$, more specific for $m=1$ one has:
\begin{equation}
C_{3}\left(\Upsilon_1,p,q\,;A_2,k_2\,; A_3,k_3 \right)=\left(\Upsilon_1 \int d\Upsilon_1\right) C_{3}\left( \{\zeta_n\},\{\Upsilon_m\},p,q\,;A_2,k_2\,; A_3,k_3 \right)\Big|_{\Upsilon,\zeta=0}
\label{eq:selectio_rules}
\end{equation}
 and after the integral over $X_0$ one gets:
  \begin{equation}
{1\over 2}\Big( \Upsilon_1{\cdot}A_2\,A_3{\cdot}{[(p{-}q/2){-}k_2]} + A_{3}{\cdot}A_2\, \Upsilon_1{\cdot}{(k_3{-}k_2)}+ \Upsilon_1{\cdot}A_3 \,A_2{\cdot}{[k_3{-}(p{-}q/2)]} \Big){(2\pi)}^D \,\delta\left(p{-}q/2{+}k_2{+}k_{3} \right)
\label{eq:selection_rules_compl}
 \end{equation}
 that is exactly the three vector coupling where one vector is identified with $\Upsilon_1$ and the respective momentum that is $p{-}q/2$. In general to select the coupling of a single state in the amplitude one can apply the operator: 
 \begin{equation}
\left( \Upsilon_{m_{*}} \int d\Upsilon_{m_{*}}\right)^{\ell_{m_{*}}}    \left(\zeta_{n_{*}} {\pa \over \pa \zeta_{n_{*}}}\right)^{g_{n_{*}}} C_{3}\left( \{\zeta_n\},\{\Upsilon_m\},p,q\,;A_2,k_2\,; A_3,k_3 \right)\Big|_{\Upsilon,\zeta=0}
\label{eq:selection_rules_gen}
 \end{equation}
for fixed values of $m_{*}$,$n_{*}$,$\ell_{m_{*}}$,$g_{n_{*}}$

\subsection{Coherent state-Vector-Vector coupling (non-canonical picture)}
 Another realization of the same three-points coupling can be obtained choosing different superghost pictures. We choose to describe the computation principally for two reasons: first as a consistency check in preparation of one loop computations on oriented and unoriented surfaces and second to show $PSL(2,\mathbb{R})$ invariance under a non trivial cancellation between sub-amplitudes, that is different with respect to the previous case in which there were only non trivial cancellations of the dependence on the coordinates of the puncture inside each single sub-amplitude. The present coupling is :
 \begin{equation}\label{3pt coupling}
C_{3}\left( \{\zeta_n\},\{\Upsilon_m\},p,q\,;A_2,k_2\,; A_3,k_3 \right)=\left\langle c{\cal V}^{NS}_{{\cal C}\,(0)}(z_1)\, c{\cal V}_{A\,(-1)}(z_2)\, c{\cal V}_{A\,(-1)}(z_3)\right\rangle
\end{equation}
Using the same setup as for the previous subsection, there are seven sub-amplitudes that read 
\begin{eqnarray}\label{S1p0}
&&S_1=\left\langle e^{{\cal O}_2}{\cal H}_{1}^{+}\,e^{ip{\cdot}X}(z_1) \,A_2{\cdot}\p e^{ik_2{\cdot}X}(z_2)\, A_3{\cdot}\p e^{ik_3{\cdot}X}(z_3)  \right\rangle \\
&&S_2=-\left\langle e^{{\cal O}_2}{\cal H}_{2}^{+}\,e^{ip{\cdot}X}(z_1) \,A_2{\cdot}\p e^{ik_2{\cdot}X}(z_2)\, A_3{\cdot}\p e^{ik_3{\cdot}X}(z_3)  \right\rangle  \\
&&S_3=\left\langle e^{{\cal O}_2}p{\cdot}\p\,{\cal H}_{1}\,e^{ip{\cdot}X}(z_1) \,A_2{\cdot}\p e^{ik_2{\cdot}X}(z_2)\, A_3{\cdot}\p e^{ik_3{\cdot}X}(z_3)  \right\rangle \\
&&S_4=\left\langle e^{{\cal O}_2}{\cal H}_1^{+}\,({\cal H}_{1})^2\,e^{ip{\cdot}X}(z_1) \,A_2{\cdot}\p e^{ik_2{\cdot}X}(z_2)\, A_3{\cdot}\p e^{ik_3{\cdot}X}(z_3)  \right\rangle \\
&&S_5=-\left\langle e^{{\cal O}_2}{\cal H}_2^{+}\,({\cal H}_{1})^2\,e^{ip{\cdot}X}(z_1) \,A_2{\cdot}\p e^{ik_2{\cdot}X}(z_2)\, A_3{\cdot}\p e^{ik_3{\cdot}X}(z_3)  \right\rangle \\
&&S_6=\left\langle e^{{\cal O}_2}{\cal O}_3^{+}\,{\cal H}_{1}\,e^{ip{\cdot}X}(z_1) \,A_2{\cdot}\p e^{ik_2{\cdot}X}(z_2)\, A_3{\cdot}\p e^{ik_3{\cdot}X}(z_3)  \right\rangle \\
&&S_7=\left\langle e^{{\cal O}_2}{\cal O}_2^{+}\,{\cal H}_{1}\,e^{ip{\cdot}X}(z_1) \,A_2{\cdot}\p e^{ik_2{\cdot}X}(z_2)\, A_3{\cdot}\p e^{ik_3{\cdot}X}(z_3)  \right\rangle 
\end{eqnarray}
The manifestly $PSL(2,\mathbb{R})$ invariant combinations consist of:
\begin{equation}
\begin{split}
&S_{A}=\left\langle c(z_1) c(z_2) c(z_3) \right\rangle\, \left\langle e^{-\phi}(z_1) e^{-\phi}(z_2)\right\rangle\,(S_1+S_2)\\
&=\int_{{\cal M}_D} d^D X_0 \,e^{F(X_0)} \sum_{m=1}{(-)^{m{+}1}\over 2^{m{+}1}} A_2{\cdot}A_3\,\widetilde{\Upsilon}_{m}{\cdot}(k_2{-}k_3)\\
&\quad \Big[  \sum_{\ell=0}  {\cal Q}^{(1/2,m)}_{m{-}\ell{-}1} \Big( 1{+}(-)^{\ell} \Big) -  \sum_{h_1,h_2=0} (-)^{h_1}{\cal Q}^{(-1/2,m)}_{m{-}h_1{-}h_{2}{-}1} \Big(1{+}(-)^{h_2}\Big)\Big((-)^{h_1}(Q{-}1){+}1{+}Q \Big)  \Big] \quad\quad\quad\quad\quad\quad
\end{split}
\label{eq:SA}
\end{equation}

\begin{equation}
\begin{split}
&S_B=\left\langle c(z_1) c(z_2) c(z_3) \right\rangle\, \left\langle e^{-\phi}(z_1) e^{-\phi}(z_2)\right\rangle\,S_3\\
&=\int_{{\cal M}_D} d^D X_0 \,e^{F(X_0)} \sum_{m=1}{(-)^{m{-}1}\over 2^{m}} \sum_{h=0}{\cal Q}_{m{-}h{-}1}^{(1/2,m)} \left( \widetilde{\Upsilon}_m{\cdot}A_2\,A_3{\cdot}(p_{{\cal C}}{-}k_2)+ (-)^{h{+}1}\widetilde{\Upsilon}_m{\cdot}A_3\,A_2{\cdot}(p_{{\cal C}}{-}k_3)\right)\quad\quad\quad
\end{split}
\label{eq:SB}
\end{equation}
\begin{equation}
\begin{split}
&S_C=\left\langle c(z_1) c(z_2) c(z_3) \right\rangle\, \left\langle e^{-\phi}(z_1) e^{-\phi}(z_2)\right\rangle\,(S_4+S_5)\\
&=\int_{{\cal M}_D} d^D X_0 \,e^{F(X_0)} \sum_{m_{1},m_{2},m_{3}=1}{(-)^{m_{1}{+}m_{2}{+}m_{3}}\over 2^{m_{1}{+}m_{2}{+}m_{3}{-}2}} \sum_{h_{2},h_{3}=0} {\cal Q}_{m_{2}{-}h_{2}{-}1}^{(1/2,m_{2})}  {\cal Q}_{m_{3}{-}h_{3}{-}1}^{(1/2,m_{3})}         \\
&\quad \Big\{ \widetilde{\Upsilon}_{m_1}{\cdot}k_2 \widetilde{\Upsilon}_{m_2}{\cdot}A_3\widetilde{\Upsilon}_{m_3}{\cdot}A_2 \Big[\sum_{h_1=0} {\cal Q}_{m_{1}{-}h_{1}{-}1}^{(1/2,m_{1})} (-)^{1{+}h_{2}}\Big( 1{+}(-)^{h_{1}} \Big) {+}\\
&+{1\over 2} \sum_{\ell_{1},\ell_{2}=0} {\cal Q}_{m_{1}{-}\ell_{1}{-}\ell_{2}{-}1}^{(-1/2,m_{1})}(-)^{\ell_{1}+h_{3}}  \Big( 1+(-)^{\ell_{2}} \Big) [(-)^{\ell_{1}}(Q-1)+Q+1] \Big] +\\
&-\widetilde{\Upsilon}_{m_1}{\cdot}k_2 \widetilde{\Upsilon}_{m_3}{\cdot}A_3\widetilde{\Upsilon}_{m_2}{\cdot}A_2 \Big[\sum_{h_1=0} {\cal Q}_{m_{1}{-}h_{1}{-}1}^{(1/2,m_{1})} (-)^{1+h_{3}}\Big( 1+(-)^{h_{1}} \Big) {+}\\
&+{1\over 2} \sum_{\ell_{1},\ell_{2}=0} {\cal Q}_{m_{1}{-}\ell_{1}{-}\ell_{2}{-}1}^{(-1/2,m_{1})}(-)^{\ell_{1}+h_{2}}  \Big( 1+(-)^{\ell_{2}} \Big) [(-)^{\ell_{1}}(Q-1)+Q+1] \Big] \Big\}
\end{split}
\label{eq:SC}
\end{equation}
\begin{equation}
\begin{split}
&S_{D}=\left\langle c(z_1) c(z_2) c(z_3) \right\rangle\, \left\langle e^{-\phi}(z_1) e^{-\phi}(z_2)\right\rangle\,(S_6+S_7)\\
&=\int_{{\cal M}_D} d^D X_0 \,e^{F(X_0)}\,    \sum_{m=1} \sum_{\ell=0} {(-)^{m{+}1} \over 2^{m}} {\cal Q}_{m{-}\ell{-}1}^{(1/2,m)}  \\
&\quad\Bigg\{\sum_{n=1}  \widetilde{\Upsilon}_{m}{\cdot}A_2\,\widetilde{\zeta}_n{\cdot}A_3 \Bigg[{\Gamma[1{+}{n\over 2}(1{-}Q)]\over n!\, \Gamma[2{-}{n\over 2}(1{+}Q)]} + {(-)^n \over 2^n} \sum_{h_1,h_2=0} {\cal Z}_{n{-}h_1{-}h_2{-}1}\,(-)^{h_1{+}h_2} \left[ (-)^{h_1}(Q{-}1) {+}Q{+}1 \right] \Bigg] +\\
&-\sum_{n=1} \widetilde{\zeta}_n{\cdot}A_2\,\widetilde{\Upsilon}_{m}{\cdot}A_3 \Bigg[{(-)^{n{+}\ell}\Gamma[1{+}{n\over 2}(1{+}Q)]\over n!\, \Gamma[2{+}{n\over 2}(Q{-}1)]} + {(-)^{n{+}h_1{+}\ell} \over 2^n} \sum_{h_1,h_2=0} {\cal Z}_{n{-}h_1{-}h_2{-}1}\,\left[ (-)^{h_1}(Q{-}1) {+}Q{+}1 \right] \Bigg] \Bigg\}
\end{split}
\label{eq:SD}
\end{equation}
One can verify that this result is exactly the same as the previous subsection. Comparing the polarisation structures and the relative polynomial, of the two different representations, one can find interesting polynomial identities, aspects that go beyond the scope of this paper.

\newpage
 \section{Extension to the Coherent vertex operator of Ramond states}
In this section we will show how to extend the NS coherent vertex operators to the Ramond sector. In particular we will discuss two different procedures. The first one is based on the use of supersymmetry transformations in the target space supersymmetry and the second one is a direct construction starting from the degenerate Ramond vacuum following the line of the previous construction in the NS sector. 
 
 \subsection{Extension using spacetime supersymmetry}
 In superstring theory, a possible way to map NS states into their super-partner states in the Ramond sector states, makes use of target space supersymmetry \cite{Friedan:1985ge}. In a ten dimensional Minkowski spacetime with $SO(1,9)$ Lorentz group, the supercharges are Majorana-Weyl spinors $Q^{SUSY}_{\alpha}$ of real dimension sixteen, and in superstring theory due to the presence of the superghost the realization of the supercharges\footnote{In the present case, in which we consider critical open superstring with only one worldsheet supersymmetry, in the ten dimensional target space there is only one supercharge.} requires a dependence on the picture number $Q^{SUSY}_{\alpha (\widetilde{q})}$. Considering a NS vertex operator of a certain mass level with picture number $\widetilde{q}_2$, one can obtain the corresponding R vertex operator by :
\begin{equation}
\xi^{\alpha} Q^{SUSY}_{\alpha (\widetilde{q}_1)} \,  {\cal V}^{(NS)}_{(\widetilde{q}_2)} (z)\big|_{\alpha'(\text{Mass})^2} = {\cal V}^{(R)}_{(\widetilde{q}_1+\widetilde{q}_2)} (z)\big|_{\alpha'(\text{Mass})^2}
\label{eq:RfromSusy}
\end{equation}

In the specific case of the NS coherent vertex operator the action of the supercharge plays exactly the same role. To compute the R coherent vertex operator in the canonical picture $q=-1/2$ one needs the supercharge in the first non canonical picture $q'=+1/2$:
 \begin{equation}
  Q^{SUSY}_{\alpha (+1/2)}=\oint {d w \over 2\pi i} \,e^{\phi/2} i\pa X_M \Gamma^{M}_{\alpha \beta} C^{\beta}(w)
\label{eq:QSUSY}
  \end{equation} 
 where $\Gamma^{M}_{\alpha \beta}$ is the ``projected" Dirac gamma matrix and $C^{\beta}$ is the spin field with opposite chirality with respect to the supercharge. To extend in the R sector the NS coherent vertex operator one can use the following relation
 \begin{equation}\label{Susy trasf NS}
\xi^{\alpha} Q^{SUSY}_{\alpha (+1/2)}\, {\cal V}_{{\cal C} (-1)}^{(NS)}(z) ={\cal V}_{{\cal C} (-1/2)}^{(R)}(z)
\end{equation} 
In order to determine the explicit structure of the Ramond coherent vertex operator one has to evaluate the OPE (\ref{Susy trasf NS}).

\subsection{Supersymmetric DDF construction}
Following the line of the construction of NS coherent vertex operator where the world-sheet supersymmetric version of the DDF operators were used, one can build a Ramond coherent vertex operator choosing the vacuum state to be the lowest state of the Ramond sector. In the light-cone  ``description'' only physical states are included and one has to consider the $SO(8)$ group associated to the transverse eight dimensions. Before imposing Majorana and Weyl conditions, the on shell spinors of $SO(1,9)$ have sixteen complex components that are the same as for the $SO(8)$ spinors before the Majorana and Weyl conditions\footnote{Majorana-Weyl spinors exist in $SO(1,9)$ and in SO(8).}. The vacuum that one can choose is a state created by an operator in the canonical picture composed by a massless Majorana spin field of $SO(8)$ having the following form 
\begin{equation}
\ket{\alpha,k}_{(-1/2)}=e^{-\phi/2} S_{\alpha} \,e^{ik{\cdot}X}(z) \,\ket{0}\,;\quad k^2=0\,,\quad \alpha=1\dots16
\label{eq:Rvacuum}
\end{equation}
Now we define the R coherent state in the canonical picture prior to the GSO projection as
\begin{equation}\label{Ramond coh state1}
\exp\left(\sum_{n=1} {\zeta_{n} \over n}{\cdot}{\cal A}_{-n} \right)\exp\left(\sum_{m=1} {\Upsilon_{m}}{\cdot}{\cal B}_{-m} \right)\ket{\alpha,k}_{(-1/2)}
\end{equation}
In terms of eight-dimensional irreducible spinorial representations of $SO(8)$ the vacuum can be decomposed in two vacua, one for the (S) conjugacy class and the other for the (C) conjugacy class. This decomposition can be realized with the use of the chirality operator $\gamma^{9}$ under which states of a definite chirality can be projected. In particular one has
\begin{equation}
P_S \ket{\alpha,k}_{(-1/2)}=\ket{a,k}^{(S)}_{(-1/2)}\,;\quad P_C \ket{\alpha,k}_{(-1/2)}=\ket{\dot{a},k}^{(C)}_{(-1/2)}\,;\quad P_{S,C}={\mathrm{1} \pm \gamma^9 \over 2}
label{eq:Rvacuums}
\end{equation}
where $P_{S,C}$ are the chirality projectors. Defining the GSO projectors in the R sector to be
 \begin{equation}
 \rho_{GSO}^{(R)\pm}={\mathrm{1} \pm \gamma^9(-)^F \over 2} \,; \quad F=\sum {\cal B}_{-m}\,{\cal B}_{m}
\label{eq:R-GSO-Proj} 
 \end{equation}
 one can split in two parts (\ref{Ramond coh state1}), giving rise to
 \begin{equation}
 ( \rho_{GSO}^{(R)+} +  \rho_{GSO}^{(R)-})\,\exp\left(\sum_{n=1} {\zeta_{n} \over n}{\cdot}{\cal A}_{-n} \right)\exp\left(\sum_{m=1} {\Upsilon_{m}}{\cdot}{\cal B}_{-m} \right)\ket{\alpha,k}_{(-1/2)}
\label{eq:R-GSOonCohR} 
 \end{equation}
 From the first term one obtains
 \begin{equation}
 \begin{split}
 &\exp\left(\sum_{n=1} {\zeta_{n} \over n}{\cdot}{\cal A}_{-n} \right) \cosh\left(\sum_{m=1} {\Upsilon_{m}}{\cdot}{\cal B}_{-m} \right)\ket{a,k}^{(S)}_{(-1/2)} + \\
 &+  \exp\left(\sum_{n=1} {\zeta_{n} \over n}{\cdot}{\cal A}_{-n} \right) \sinh\left(\sum_{m=1} {\Upsilon_{m}}{\cdot}{\cal B}_{-m} \right)\ket{\dot{a},k}^{(C)}_{(-1/2)}
 \end{split}
 \label{eq:R-GSOonCohR1}
 \end{equation}
that is the combination with $+1$ eigenvalue under the action of the operator $\gamma^9 \,(-)^F$, while from the second term one has the combination with $-1$ eigenvalue, that is
 \begin{equation}
 \begin{split}
 &\exp\left(\sum_{n=1} {\zeta_{n} \over n}{\cdot}{\cal A}_{-n} \right) \cosh\left(\sum_{m=1} {\Upsilon_{m}}{\cdot}{\cal B}_{-m} \right)\ket{\dot{a},k}^{(C)}_{(-1/2)} + \\
 &+  \exp\left(\sum_{n=1} {\zeta_{n} \over n}{\cdot}{\cal A}_{-n} \right) \sinh\left(\sum_{m=1} {\Upsilon_{m}}{\cdot}{\cal B}_{-m} \right)\ket{a,k}^{(S)}_{(-1/2)}
 \end{split}
 \label{eq:R-GSOonCohR2}
 \end{equation}
 The GSO projection in the Ramond sector requires to choose only states with the same eigenvalue of $\gamma^9 \,(-)^F$, as a result our definition of GSO projected Ramond coherent vertex operator in the canonical picture is
 
 \begin{equation}\label{Ramond cohe finale}
 {\cal V}^{(R)}_{{\cal C}(-1/2)}(z)=\rho_{GSO}^{(R)+}\exp\left(\sum_{n=1} {\zeta_{n} \over n}{\cdot}{\cal A}_{-n} \right)\exp\left(\sum_{m=1} {\Upsilon_{m}}{\cdot}{\cal B}_{-m} \right)\,e^{-\phi/2}\,S_{\alpha}\,e^{ik{\cdot}X}(z)
 \end{equation}

It would be very interesting to determine the normal ordered version of (\ref{Ramond cohe finale}) as we did for the NS coherent vertex operator. We leave it to the future.

\section{Summary, conclusion and outlook} 
In the context of open superstring theory, we have used the SGA of the transverse supersymmetric version of the DDF operators to construct arbitrarily massive NS higher-spin physical (BRST invariant) states in the canonical picture, where without loss of generality only a single D9 brane was considered, providing a general formula in a closed form. To determine the structure of the arbitrary massive higher-spin vertex operator is necessary to perform the normal ordering between the supersymmetric DDF operators and the operator representing the selected vacuum. We have classified the OPE rules required for the derivation.

We have then constructed the NS coherent vertex operator in the canonical picture in terms of DDF operators and the different behaviour between Grassmann even and Grassmann odd operators was pointed out. In fact the Grassmann even part of the state turns out to be coherent and identified with the parameters (quantum numbers) $\zeta^{\mu}_n$ that are continuous c-numbers. The Grassmann odd part of the state is coherent and sqeezed with the parameters represented by $\Upsilon^{\mu}_m$ that are continuous a-numbers. After the required normal ordering we have eliminated the explicit dependence on the DDF operators and found a new exponentiation with a set of operatorial structures constrained by the GSO projection. Owing to the dictionary between the mass shell and the new operators insertions we have studied the first few mass levels (as a level expansion of the coherent vertex operator) finding, level by level, the vertex operators with the counting of the d.o.f. in agreement with the literature. This allows us to conclude that the normal ordered coherent vertex operator includes the whole content of the NS sector being a generating function of BRST invariant and manifestly covariant vertex operators. 

Motivated by the interest of computing scattering amplitudes involving more than one NS coherent vertex operator, we have derived the vertex operator in the first non canonical superghost picture which is also necessary for the computation of the first higher genus i.e. one loop string scattering amplitudes. As a consistency check we have computed the coupling of one NS coherent state, in both superghost pictures and two vectors finding complete agreement among them. We have given the rules to select the coupling at the desired level in the coherent superposition.

To complete the revelation of coherent states in open superstring theory, we have illustrated two different procedures for extending our construction to the Ramond sector. The first is based on the use of supersymmetric transformations and  the second on a direct construction carried out with the help of the DDF operators. In this last case the identification of the GSO projections is required to obtain a supersymmetric theory in the target space with only physical states.

The generalization to the closed superstring theory is not too difficult. One has to introduce the level matching between left and right modes through a delta integral with a dressing of the polarisations and compose the GSO projections of the Ramond sector to recover the closed string sectors. In the case of type \RNum{2}A and type \RNum{2}B theories the NS-NS coherent vertex operator in the canonical picture takes the form
\begin{equation}
\begin{split}
{\cal W}^{(NS)(NS)}_{{\cal C}(-1)(-1)}(z,\bar{z})=\int_{0}^{2\pi} d\beta\, &\exp\left(\sum_{n=1}^{\infty}{1\over n} {\zeta^{(\beta)}_{n} }{\cdot}{\cal A}^{(L)}_{-n} \right) \, \rho^{(NS)+}_{GSO,L}\,\exp\left(\sum_{m=1}^{\infty} {\Upsilon^{(\beta)}_m }{\cdot}{\cal B}^{(L)}_{-m} \right)\,e^{-\phi}e^{ip{\cdot}X_L}(z) \\
&\exp\left(\sum_{\bar{n}=1}^{\infty}{1\over \bar{n}} {\bar{\zeta}^{(\beta)}_{\bar{n}} }{\cdot}{\cal A}^{(R)}_{-\bar{n}} \right)\, \rho^{(NS)+}_{GSO,R}\, \exp\left(\sum_{\bar{m}=1}^{\infty} {\bar{\Upsilon}^{(\beta)}_{\bar{m}} }{\cdot}{\cal B}^{(R)}_{-m} \right)\,e^{-\bar{\phi}}e^{ip{\cdot}X_R}(\bar{z})
\end{split}
\label{eq:closedCohNS}
\end{equation}
with $\zeta^{\mu(\beta)}_{n}=\zeta^{\mu}_{n} e^{-in\beta}$, $\bar{\zeta}^{\mu(\beta)}_{\bar{n}}=\zeta^{\mu}_{\bar{n}} e^{i\bar{n}\beta}$ and $\Upsilon^{\mu(\beta)}_{m}=\Upsilon^{\mu}_{m} e^{-im\beta}$, $\bar{\Upsilon}^{\mu(\beta)}_{\bar{m}}=\bar{\Upsilon}^{\mu}_{\bar{m}} e^{i\bar{m}\beta}$. 

The normal ordered version is made of
\begin{equation}
{\cal W}^{(NS)(NS)}_{{\cal C}(-1)(-1)}(z,\bar{z})=\int_{0}^{2\pi} d\beta\, e^{-\phi}e^{{\cal O}^{(\beta)}_2{+}{\cal O}^{(\beta)}_3}\sinh\left[{\cal H}_{1}^{(\beta)}{-} {\cal H}_{2}^{(\beta)}\right]\,e^{-\bar{\phi}}e^{\bar{{\cal O}}^{(\beta)}_2{+}\bar{{\cal O}}^{(\beta)}_3}\sinh\left[\bar{{\cal H}}_{1}^{(\beta)}{-} \bar{{\cal H}}_{2}^{(\beta)}\right]e^{ip{\cdot}X_{L,R}}
\label{eq:closedCohNS_NO}
\end{equation}
with $X_{L,R}=X_L+X_R$ and the notation ${\cal O}_{2,3}^{(\beta)}= {\cal O}_{2,3}(\zeta^{(\beta)})$, ${\cal H}_{1,2}^{(\beta)}= {\cal H}_{1,2}(\Upsilon^{(\beta)})$.
Following the same line one can write down the coherent vertex operator for the other sectors and in particular the coherent vertex operator of the RR sector in the canonical superghost pictures following (\ref{Ramond cohe finale}) is given by
\begin{equation}
\begin{split}
&{\cal W}^{(R)(R)}_{{\cal C}(-{1\over 2})(-{1\over 2})}(z,\bar{z})=\int_{0}^{2\pi} d\beta \,{\cal V}^{(R+)}_{{\cal C}(-{1\over 2}),L}(z) \, {\cal V}^{(R+)}_{{\cal C}(-{1\over 2}),R}(\bar{z})\,;\quad \quad \text{type \RNum{2}B} \\
&{\cal W}^{(R)(R)}_{{\cal C}(-{1\over 2})(-{1\over 2})}(z,\bar{z})=\int_{0}^{2\pi} d\beta \,{\cal V}^{(R+)}_{{\cal C}(-{1\over 2}),L}(z) \, {\cal V}^{(R-)}_{{\cal C}(-{1\over 2}),R}(\bar{z})\,;\quad \quad \text{type \RNum{2}A} 
\end{split}
\label{eq:closedCohR_A/B}
\end{equation}
 where the superscript $(R+)$ and $(R-)$ of the holomorphic and antiholomorphic vertex operators are referred to the choice of GSO projections $\rho^{(R)+}_{GSO}$ and  $\rho^{(R)-}_{GSO}$, respectively.
 Of course to be able to cover many different deep aspects of the superstring theory one needs to extend the present construction to the case of asymmetric pictures, as pointed out in \cite{Bianchi:1991eu}, and in the case of open superstring the inclusion of the Chan-Paton factors is needed to consider multiple branes setups. We argue that this construction could be a good starting point for the study of the interactions of higher spin states \cite{Chialva:2014rla,Chialva:2006ak} and macroscopic states \cite{Jackson:2007hn,Lidsey:1999mc} in superstring theory. In addition, it would be very interesting to extend the analysis in the context of strings living in Anti-de Sitter spaces \cite{Maldacena:1997re, Itzhaki:1998dd, Berenstein:2002jq}.


\vspace{3mm}
{\large \bf Acknowledgments}

We would like to thank Massimo Bianchi and Giancarlo Rossi for enlightening discussions and comments about the manuscript and  Dimitri Skliros for having motivated our interest in the subject.

\newpage 
\begin{appendix}
\section{Polynomial structures} 
The first non trivial polynomial structure is the cycle index polynomial of the symmetric group
 \begin{equation}
{\cal Z}_{N}[{U}^{(\ell)}_{k}]:={ \pa_{z}^{N}\over N!} \,\exp\left(\sum_{k=1}^{\infty} {(-\ell)\,{U}_{k} \over k} z^k\right)\Big|_{z=0}
\label{eq:Zpoly}
 \end{equation}
the second polynomial structure that one encounters is ${\cal G}$:
\begin{equation}
\begin{split}
{\cal G}_{N}^{(1/2)}[U_k^{(-1)}]&:={\pa_{z}^{N}\over N!}  \left(1 + \sum_{s=1} { U}^{(-1)}_{s}\,z^{s}  \right)^{1/2} \Bigg|_{z=0}\\
&= {1\over \Gamma\left(-{1\over 2}\right)}\int_{0}^{\infty} d\eta\, \eta^{-{1\over 2}-1} e^{-\eta} \,{\pa_{z}^{N} \over N!}\, \exp\left(-\eta \sum_{s=1}{ U}_{s}^{(-1)}\,z^s \right)\Bigg|_{z=0}\\
&= {1\over \Gamma\left(-{1\over 2}\right)}\int_{0}^{\infty} d\eta\, \eta^{-{1\over 2}-1} e^{-\eta} \,{\cal Z}_{N}\left[ s\,{U}_{s}^{(\eta)}\right]\\
&={1\over \Gamma\left(-{1\over 2}\right)}\int_{0}^{\infty} d\eta\, \eta^{-{1\over 2}-1} e^{-\eta} \sum_{\ell_{s}:\sum_{s}s \ell_{s}=N} \prod_{s=1}^{N} \left({  \eta\, s\,{U}^{(1)}_{s}  \over s}\right)^{\ell_{s}} {1\over \ell_{s}!}\\
&={1\over \Gamma\left(-{1\over 2}\right)}\,\sum_{\ell_{s}:\sum_{s}s\, \ell_{s}=N}\, \Gamma\left( \sum_{s}\ell_s -\dfrac{1}{2} \right) \prod_{s=1}^{N} \left( {s\,{U}^{(1)}_{s}\over s} \right)^{\ell_s} {1\over l_s !}
\end{split}
\label{eq:Gpoly}
\end{equation}
where the general formula is represented by
\begin{equation}
{\cal G}^{(k)}_{N}[{U}^{(-1)}_{s}]:=  {1\over \Gamma\left(-k\right)}\,\sum_{\ell_{s}:\sum_{s}s \ell_{s}=N}\, \Gamma\left( \sum_{s}\ell_s -k\right) \prod_{s=1}^{N} \left( {s\,{U}^{(1)}_{s}\over s} \right)^{\ell_s} {1\over l_s !}
\label{eq:Gpoly2}
\end{equation}
for instance one can give some low degree expressions of this polynomial:
\begin{eqnarray}
&&{\cal G}^{(k)}_{-N}[a_s]=0\,;\quad {\cal G}^{(k)}_{0}[a_s]=1 \nonumber\\
&&{\cal G}^{(1/2)}_{1}[a_s]={a_1 \over 2}\,;\quad {\cal G}^{(-1/2)}_{1}[a_s]=-{a_1 \over 2}\,;\quad {\cal G}^{(-3/2)}_{1}[a_s]=-{3 a_1 \over 2} \nonumber\\
&&{\cal G}^{(1/2)}_{2}[a_s]={1\over 2} \left(-{a_1^2 \over 4} {+}a_2\right)\,;         \quad {\cal G}^{(-1/2)}_{2}[a_s]={1\over 2} \left({3 a_1^2 \over 4} {-}a_2\right)\,;         \quad {\cal G}^{(-3/2)}_{2}[a_s]={1\over 2} \left({15 a_1^2 \over 4} {-}3a_2\right)\nonumber\\\nonumber
\label{eq:Gpoly_ex}
\end{eqnarray}
The last polynomial structure is a combination of the last two polynomials and has the following form
\begin{equation}
\begin{split}
 { \cal Q}^{(k,m)}_{N}[{U_s}^{({-}1)};{U_{\ell}}^{(m{-}1/2)}]&:={\pa_{z}^{N} \over N! } \left\{   \left(1 + \sum_{s=1} {U}^{(-1)}_{s} z^{s}  \right)^{k}  \exp\left(\sum_{\ell=1} {{U}_{\ell} \over \ell}^{(m{-}1/2)} z^{\ell}\right) \right\}\Bigg|_{z=0}\\
 &\,= \sum_{v=0}^{N} {\cal G}^{(k)}_{v}[U_{\ell}^{(-1)}] \,{\cal Z}_{N-v}[{U}_s^{(m-1/2)}]
 \end{split}
 \label{eq:Qpoly}
\end{equation}
In the computation of the discussed scattering amplitudes there are many sums that appear and here we collect some useful resummation formulae\footnote{We will not indicate the upper bounds of the summations because they are implicitly limitated by the presence of the polynomial structures $\cal Z$ and $\cal Q$, indeed ${\cal Z}_{-\ell}={\cal Q}_{-\ell}=0$.}. 
\begin{itemize}
\item Sums involving cycle index polynomial that appear in the first amplitude (NS coherent operator in canonical superghost picture) of the section 2
\end{itemize}
 \begin{equation}
 \sum_{h=1}{(-)\over 2 n}^{h{+}1} {\cal Z}_{n{-}h}\left[{\cal U}^{(n)}_{s}\right] \left(z_{12}^{-h} - z_{13}^{-h} \right) \left( {z_{12}z_{13} \over z_{23}} \right)^n = {(-)^{n{+}1}\over 2 \,n!} {\Gamma\left[ {n\over 2}(Q{+}1) \right] \over \Gamma\left[{n\over 2}(Q{-}1){+}1) \right]}
\label{eq:resumZ1}
 \end{equation}

 \begin{equation}
 \sum_{h_1,h_2=0}{(-)^{h_{1}{+}h_{2}} \over z_{13}^{h_{1}{+}h_{2}}} {\cal Z}_{n{-}h_1{-}h_2{-}1}\left[{\cal U}^{(n)}_{s}\right]  \left( {z_{12}z_{13} \over z_{23}} \right)^n = \left( {z_{12}z_{13} \over z_{23}} \right) {1\over \,(n{-}1)!} {\Gamma\left[ {n\over 2}(1{-}Q){+}1 \right] \over \Gamma\left[2{-}{n\over 2}(Q{+}1)) \right]}
 \label{eq:resumZ2}
 \end{equation}

  \begin{equation}
 \sum_{h_1,h_2=0}{(-)^{h_{1}{+}h_{2}} \over z_{13}^{h_{1}} z_{12}^{h_2}} {\cal Z}_{n{-}h_1{-}h_2{-}1}\left[{\cal U}^{(n)}_{s}\right]  \left( {z_{12}z_{13} \over z_{23}} \right)^n = \left( {z_{12}z_{13} \over z_{23}} \right) {1\over \,(n{-}1)!} {\Gamma\left[ {n\over 2}(1{-}Q) \right] \over \Gamma\left[1{-}{n\over 2}(Q{+}1)) \right]}
 \label{eq:resumZ3}
 \end{equation}
 
 where after the contractions the operatorial argument of the cycle index polynomials becomes:
 \begin{equation}
 {\cal U}_{s}^{(n)}=(-)^s {n\over 2}\left( {Q-1 \over z_{12}^s} - {Q+1 \over z_{13}^s} \right) \,; \quad Q=q{\cdot}(p_2-p_3)
 \label{eq:Udef}
 \end{equation}
 
 \begin{itemize}
\item Sums involving cycle index polynomial that appear in the last amplitude (NS coherent operator in the first non canonical superghost picture) of the section 2
\end{itemize}

\begin{equation}
\sum_{h=1} (-)^h\, h \,{\cal Z}_{n{-}h} \,z_{13}^{-h}  \left( {z_{12}z_{13} \over z_{23}} \right)^n=-{z_{12}\over z_{23}}  {1 \over (n{-}1)!} {\Gamma[1{+}{n\over 2}(1{-}Q)] \over \Gamma[2{-}{n\over 2}(1{+}Q)]} 
\label{eq:resumZ4}
\end{equation}
\begin{equation}
\sum_{h=1} (-)^h\, h \,{\cal Z}_{n{-}h} \,z_{12}^{-h}  \left( {z_{12}z_{13} \over z_{23}} \right)^n={z_{13}\over z_{23}}  {(-1)^n \over (n{-}1)!} {\Gamma[1{+}{n\over 2}(Q{+}1)] \over \Gamma[2{+}{n\over 2}(Q{-}1)]} 
\label{eq:resumZ5}
\end{equation}
 
\section{Classification of operators}
The following list displays the classification of the operators that appear in constuction of arbitrary massive NS higher-spin vertex operators and the NS coherent vertex operators both in canonical picture. The classification is separately done according to the DDF action by which they are generated.
From the ${\cal A}$ actions one gets
\begin{eqnarray}
 &&{\cal S}_{m,n}= \sum_{h=1} h \, {\cal Z}_{m+h}[{\cal U}^{(m)}_s] {\cal Z}_{n-h}[{\cal U}^{(n)}_s]\,;\quad {\cal P}^{i}_n=\sum_{h=1} {\cal Z}_{n-h}[{\cal U}^{(n)}_s] {i \pa^h X^{i} \over n (h-1)!}\\
 &&{\cal F}_{n-1,m}=\sum_{h_{1},h_{2},h_{3}} (-)^{h_{1}+1}  {\cal Z}_{n{-}1{-}h_{1}{-}h_{2}}[{\cal U}^{(n)}_s]  {\cal Z}_{m{+}h_{1}{-}h_{3}}[{\cal U}^{(m)}_s]\,{q{\cdot}\pa^{h_{2}}\p \over h_{2}!}\, {q{\cdot}\pa^{h_{3}}\p \over h_{3}!}\\
 &&{\cal E}^{i}_{n-1}=-\sum_{h_{1},h_{2}=0}  {\cal Z}_{n{-}1{-}h_{1}{-}h_{2}}[{\cal U}^{(n)}_s]  \,{q{\cdot}\pa^{h_{1}}\p \over h_{1}!} \,{\pa^{h_{2}}\p^{i} \over h_{2}!}
 \label{eq:operators_A}
  \end{eqnarray} 
where we have maintained the explicit dependence of the argument of the ${\cal Z}$ polynomial, that we will omitt in the next terms being always the same, and the same   for the ${\cal Q}$ polynomial. From the combined actions
 \begin{eqnarray} 
&&{\cal Y}^{({1\over 2})}_{n{-}1,m}=\hspace*{-0.4cm}\sum_{h_1,h_2=0}q{\cdot}{\pa^{h_1}\p \over h_1 !}\, {\cal Z}_{n{-}h_{1}{-}h_{2}{-}1} \, { {\cal Q}}^{(1/2,m)}_{m{+}h_{2}}\,;\quad  {\cal Y}^{(-{1\over 2})}_{n,m{-}1}=\hspace*{-0.4 cm}\sum_{h_1,h_2=0}h_1\,q{\cdot}{\pa^{h_2}\p \over h_2 !}\, {\cal Z}_{n{-}h_{1}} \, { {\cal Q}}^{(1/2,m)}_{m{+}h_{1}{-}h_2{-}1} \nonumber\\
\\
&&{\cal Y}^{(-{3\over 2})}_{n{-}1,m{-}1}=\sum_{h_1,h_2,h_{3},h_{4}=0}q{\cdot}{\pa^{h_1}\p \over h_1 !} q{\cdot}{\pa^{h_3}\p \over h_3 !}  q{\cdot}{\pa^{h_4}\p \over (h_4 {-}1) !}\, {\cal Z}_{n{-}h_{1}{-}h_{2}{-}1} \, { {\cal Q}}^{(-3/2,m)}_{m{+}h_{2}{-}h_3{-}h_{4}{-}1}
\label{eq:operators_AB}
 \end{eqnarray} 
From the ${\cal B}$ actions
\begin{eqnarray}
&&{\cal O}^{i,\p}_{n{-}1}=\sum_{h=0} {\pa^h \p^i \over h!} \,{ {\cal Q}}^{(1/2,n)}_{n{-}1{-}h}\,;\quad {\cal O}^{i,\p\p\p}_{n{-}2}=\hspace*{-0.5cm}\sum_{h_{1},h_{2},h_{3}=0} {\pa^{h_{1}}\p^i \over 2 (h_{1}!)}\,q{\cdot}{\pa^{h_{2}}\p\over h_{2}!}\, {q{\cdot}\pa^{h_{3}}\p\over (h_{3}-1)!}\,{{\cal Q}}^{(-3/2,n)}_{n{-}2{-}h_{1}{-}h_{2}{-}h_{3}}\\
&&{\cal O}^{i,\p p}_{n{-}1}= \sum_{h_{1}=0} q{\cdot}{\pa^{h_{1}}\p\over h_{1}!} \,p^i\, \,{ {\cal Q}}^{(-1/2,n)}_{n{-}1{-}h_{1}}\,;\quad {\cal O}^{i,\p \pa X}_{n{-}1}=\sum_{h_{1}=0} q{\cdot}{\pa^{h_{1}}\p\over h_{1}!} \,\sum_{h_{2}=1}{i\pa^{h_{2}}X^i \over (h_{2}{-}1)!}\,\,{ {\cal Q}}^{(-1/2,n)}_{n{-}1{-}h_{1}{-}h_{2}}\\
&&{\cal K}^{(-\frac{1}{2}),(-\frac{1}{2})}_{n{-}1,m{-}1}=\sum_{h_{1},h_{2},h_{3}} q{\cdot}{\pa^{h_{1}}\p\over h_{1}!}\,q{\cdot}{\pa^{h_{3}}\p\over h_{3}!} \,h_{2}\,{{\cal Q}}^{(-1/2,n)}_{n{-}1{-}h_{1}{-}h_{2}} \,\,{ {\cal Q}}^{(-1/2,m)}_{m{-}1{-}h_{3}{+}h_{2}}\\
 &&{\cal K}^{(-\frac{3}{2}),(\frac{1}{2})}_{n{-}2,m}={1\over 2}\sum_{h_{1},h_{2},h_{3}=0}  (-)^{h_{1}{+}1} q{\cdot}{\pa^{h_{2}}\p \over h_{2}!}\,{q{\cdot}\pa^{h_{3}}\p \over (h_{3}{-}1)!}\,\,{ {\cal Q}}^{(-3/2,n)}_{n{-}2{-}h_{2}{-}h_{3}{-}h_{1}}  \,{ {\cal Q}}^{(1/2,m)}_{m{+}h_{1}} \\
 &&{\cal K}^{(\frac{1}{2}),(-\frac{3}{2})}_{n{-}1,m{-}1}={1\over 2}\sum_{h_{1},h_{2},h_{3}=0}(-)^{h_{1}{+}1} q{\cdot}{\pa^{h_{2}}\p \over h_{2}!}\,{q{\cdot}\pa^{h_{3}}\p \over (h_{3}{-}1)!}\,\,{ {\cal Q}}^{(1/2,n)}_{n{-}1{-}h_{1}}\,{ {\cal Q}}^{(-3/2,m)}_{m{-}1{-}h_{2}{-}h_{3}{+}h_{1}} \\
&&{\cal K}^{(-\frac{3}{2}),(-\frac{3}{2})}_{n{-}2,m}=\hspace*{-0.5cm}\sum_{\hspace*{-0.5 cm}h_{1},h_{2},h_{3},\ell_{1},\ell_2 =0} \hspace*{-0.2 cm} {(-)\over 4}^{h_{1}{+}1} \hspace*{-0.4 cm} q{\cdot}{\pa^{h_{2}}\p \over h_{2}!}{q{\cdot}\pa^{h_{3}}\p \over (h_{3}{-}1)!}     q{\cdot}{\pa^{\ell_{1}}\p \over  \ell_{1}!}{q{\cdot}\pa^{l_{2}}\p \over (\ell_{2}{-}1)!}          { {\cal Q}}^{(-3/2,n)}_{n{-}2{-}h_{2}{-}h_{3}{-}h_{1}} { {\cal Q}}^{(-3/2,m)}_{m{-} \ell_{1}{-}\ell_{2}{+}h_{1}} \\
&& {\cal K}^{(\frac{1}{2}),(\frac{1}{2})}_{n{-}1,m}=\sum_{h=0}(-)^{h{+}1}\,{ {\cal Q}}^{(1/2,n)}_{n{-}1{-}h} \,{ {\cal Q}}^{(1/2,m)}_{m{+}h}
\label{eq:operators_B}
 \end{eqnarray}

 Next operators are those that appear in the picture changing OPE 
 \begin{equation}
 \begin{split}
 {\cal O}_2^{+}=\sum_{n=1}^{\infty}\sum_{h=1}^{n} \left( {\zeta_{n}{\cdot}\pa^{h}\p \over n(h{-}1)!} {\cal Z}_{n{-}h} - {\zeta_{n}{\cdot}i\pa^{h} X \over (h{-}1)!} \sum_{v=0}^{n{-}h}q{\cdot}{\pa^v \p \over v!} {\cal Z}_{n{-}h{-}v}\right)\,e^{-inq{\cdot}X}
 \end{split}
 \label{O2+}
 \end{equation}

 \begin{equation}
 \begin{split}
 {\cal O}_3^{+}=\sum_{n=1}^{\infty}\sum_{h_{1}=0}^{n}\sum_{h_2=0}^{n{-}h_{1}} &\Bigg( q{\cdot}{\pa^{h_1}\p \over h_1!} \,\zeta_{n}{\cdot}{i\pa^{h_2{+}1}X \over h_2! } \,{\cal Z}_{n{-}h_1{-}h_2{-}1}-\zeta_{n}{\cdot}{\pa^{h_1}\p \over h_1!} \,q{\cdot}{i\pa^{h_2{+}1}X \over h_2! } \,{\cal Z}_{n{-}h_1{-}h_2{-}1}\,+\\
 &+n\,q{\cdot}{\pa^{h_{1}}\p\over h_1!}\, \zeta_n{\cdot}{\pa^{h_{2}}\p\over h_2!}\sum_{v=0}^{n{-}h_1{-}h_2} q{\cdot}{\pa^v\p \over v!} {\cal Z}_{n{-}h_1{-}h_2{-}v{-}1}\Bigg) e^{-inq{\cdot}X}
 \end{split}
 \label{eq:O3+}
 \end{equation}
 
 \begin{equation}
 \begin{split}
 {\cal H}_2^{+}=&\sum_{m=1}\sum_{h_1,h_2=0}\Bigg[ q{\cdot}{\pa^{h_1}\p \over h_1 !} \,{{\Upsilon}_m{\cdot}\pa^{h_2}\p \over (h_2{-}1)!} {\cal Q}^{(-1/2,m)}_{m{-}1{-}h_1{-}h_2} {+} q{\cdot}{i\pa^{h_1{+}1}X \over h_1 !} \,{{\Upsilon}_m{\cdot}i\pa^{h_2}X \over m(h_2{-}1)!} {\cal Q}^{(-1/2,m)}_{m{-}1{-}h_1{-}h_2}\,+\\
 &+q{\cdot}{\pa^{h_1}\p \over h_1!} \,{{\Upsilon}_m{\cdot}i\pa^{h_2}X \over (h_2{-}1)!} \sum_{v=0} q{\cdot}{\pa^v \p \over v!}  \left( (m{-}{1\over 2}){\cal Q}^{(-1/2,m)}_{m{-}1{-}h_1{-}h_2{-}v} + {v \over 2}{\cal Q}^{(-3/2,m)}_{m{-}1{-}h_1{-}h_2{-}v} \right) \Bigg]\,e^{-i(m{-}1/2)q{\cdot}X}
 \end{split}
 \label{H2+}
 \end{equation}
 
 \begin{equation}
 \begin{split}
  {\cal H}_1^{+}=&\sum_{m=1}\sum_{h=0}\Bigg[ \Upsilon_m{\cdot}{\pa^h\p \over\,h!}\sum_{v=0}q{\cdot}{\pa^v\p \over v!} \left( (m{-}{1\over 2}){\cal Q}^{(1/2,m)}_{m{-}1{-}h{-}v} {-}{v\over 2}{\cal Q}^{(-1/2,m)}_{m{-}1{-}h{-}v}\right) + \\
  &\hspace*{1,2cm}+ \Upsilon_m{\cdot}{i\pa^{h{+}1}X \over \,h!} {\cal Q}^{(1/2,m)}_{m{-}h{-}1}\Bigg] e^{-i(m{-}1/2)q{\cdot}X}
 \end{split}
 \label{H1+}
 \end{equation}
\end{appendix}


\end{document}